\documentclass[letterpaper, aps, prd, twocolumn, superscriptaddress, showpacs, nofootinbib]{revtex4}
\pdfoutput=1

\usepackage{graphicx}
\usepackage{bm}
\usepackage{amssymb}
\usepackage{amsmath}
\usepackage{amsfonts}
\usepackage{dcolumn}
\usepackage{color}
\usepackage[sort&compress]{natbib}
\usepackage[latin9]{inputenc}
\usepackage{url}
\usepackage{subfigure}
\usepackage{float}
\usepackage{longtable}
\sloppypar

\newcommand{\raiseentry}[1]{\smash{\raise 0.7 em \hbox{#1}}}

\newcommand{\abs}[1]{\left| {#1} \right|}
\newcommand{\cat}[1]{{\tt {#1}}}

\def\aj{Astron.\ J.}
\def\apj{Astrophys.\ J.}
\def\apjl{Astrophys.\ J. Lett.}

\def\aap{Astron.\ Astrophys.}
\def\araa{Ann.\ Rev.\ Astron.\ Astroph.}
\def\physrep{Phys.\ Rep.}
\def\mnras{Mon.\ Not.\ Roy.\ Astron.\ Soc.}

\def\prl{Phys.\ Rev.\ Lett.}
\def\prd{Phys.\ Rev.\ D}

\def\cqg{Class.\ Quantum Grav.}

\newenvironment{equationarray*}
{\arraycolsep 0.14 em
\begin{eqnarray*}}
{\end{eqnarray*}}

\begin{document}

\title{Inferring Core-Collapse Supernova Physics with Gravitational Waves}

\author{J. Logue} \email{j.logue@physics.gla.ac.uk}
\affiliation{SUPA, Institute for Gravitational Research, School of Physics and Astronomy, University of
  Glasgow, Glasgow G12 8QQ Scotland, United Kingdom} \affiliation{LIGO
  Laboratory, California Institute of Technology, Pasadena, CA 91125, USA}

\author{C. D.\ Ott}
\email{cott@tapir.caltech.edu}
\affiliation{TAPIR, MC 350-17, California Institute of Technology, Pasadena, CA 91125, USA}
\affiliation{Institute for the Physics and Mathematics of the Universe (IPMU), The University of Tokyo, Kashiwa, Japan}
\affiliation{Center for Computation and Technology, Louisiana State University, Baton Rouge, LA, USA}

\author{I. S. Heng} \email{Ik.Heng@glasgow.ac.uk}
\affiliation{SUPA, Institute for Gravitational Research, School of Physics and Astronomy, University of
  Glasgow, Glasgow G12 8QQ Scotland, United Kingdom}

\author{P. Kalmus}
\email{kalmus@caltech.edu}
\affiliation{LIGO Laboratory, California Institute of Technology, Pasadena, CA 91125, USA}
\affiliation{TAPIR, MC 350-17, California Institute of Technology,
Pasadena, CA 91125, USA}

\author{J. H. C. Scargill}
\email{james.scargill@new.ox.ac.uk}
\affiliation{New College, Oxford, OX1 3BN, UK}
\affiliation{LIGO Laboratory, California Institute of Technology, Pasadena, CA 91125, USA}

\date{\today}


\begin{abstract} 
Stellar collapse and the subsequent development of a core-collapse
supernova explosion emit bursts of gravitational waves (GWs) that
might be detected by the advanced generation of laser interferometer
gravitational-wave observatories such as Advanced LIGO, Advanced
Virgo, and LCGT. GW bursts from core-collapse supernovae encode
information on the intricate multi-dimensional dynamics at work at the
core of a dying massive star and may provide direct evidence for the
yet uncertain mechanism driving supernovae in massive stars.  
Recent multi-dimensional simulations of core-collapse supernovae
exploding via the neutrino, magnetorotational, and acoustic explosion
mechanisms have predicted GW signals which have distinct structure in
both the time and frequency domains.
Motivated by this, 
we describe a promising method for determining the most likely explosion
mechanism underlying a hypothetical GW signal, based on Principal
Component Analysis and Bayesian model selection. Using simulated
Advanced LIGO noise and assuming a single detector and linear waveform
polarization for simplicity, we demonstrate that our method can
distinguish magnetorotational explosions throughout the Milky
Way ($D \lesssim 10\,\mathrm{kpc}$) and explosions driven by the
neutrino and acoustic mechanisms to $D \lesssim
2\,\mathrm{kpc}$. Furthermore, we show that we can differentiate
between models for rotating accretion-induced collapse of massive
white dwarfs and models of rotating iron core collapse with high
reliability out to several $\mathrm{kpc}$.
\end{abstract}


\maketitle


\section{Introduction}
\label{section:introduction}

Almost eighty years after the proposal by Baade \& Zwicky that
(core-collapse) supernovae represent the transition of an ordinary
massive star into a neutron star \cite{baade:34b}, we still
lack a complete understanding of this phenomenon. In particular, we do
not know with certitude how the \emph{supernova mechanism} operates and
converts the necessary fraction of gravitational energy of collapse
into kinetic energy and light of the explosive outflow.

The basic story line of core collapse goes as follows (see
\cite{bethe:90,janka:07} for detailed reviews): At the end of a
massive star's ($8-10\,M_\odot \lesssim M \lesssim 130\,M_\odot$ at
zero-age main sequence [ZAMS]) life, nuclear burning has ceased in its
core, which is then composed primarily of iron-group nuclei (or O-Ne
nuclei at the lower end of the mass range) and supported against
gravity by the pressure of relativistically degenerate
electrons. Eventually, the core exceeds its effective Chandrasekhar
mass and dynamical collapse sets in. The collapsing core separates
into subsonically infalling homologous $(v \propto r)$ inner core and
supersonically collapsing outer core
\cite{goldreich:80,yahil:83}. When the inner core reaches nuclear
density, the repulsive core of the nuclear force leads to a stiffening
of the nuclear equation of state (EOS). The inner core, suddenly
supported by the stiff supernuclear EOS, overshoots its new
equilibrium, then bounces back into the still infalling outer core. A
shock wave forms at the sonic point between inner and outer core at an
enclosed baryonic mass of $\sim$$0.5\,M_\odot$. It quickly moves out in
radius and mass, but must do work in breaking up infalling iron-group
nuclei. This and neutrino losses from electron capture in the region
behind the shock sap its might. The shock succumbs to the ram pressure
of the outer core, stalls, and turns into an accretion shock.

The shock must be re-energized to drive a core-collapse supernova
explosion and, in the canonical scenario, leave behind a 
neutrino-cooling and contracting protoneutron star.  This shock
revival must, depending on progenitor star structure, occur within
$\sim$$0.5 - 3\,\mathrm{s}$, otherwise accretion will push the
protoneutron star over its maximum mass, leading to collapse and black
hole formation \cite{oconnor:11}.  Understanding the supernova
mechanism, which must robustly revive the stalled shock in supernovae
from massive stars that are observed on a daily basis, is the
principle current challenge of core-collapse supernova theory.

Observational clues for the supernova mechanism are few.
Electromagnetic waves are emitted in optically thin regions far from
the core and thus yield only second-hand information about the
supernova mechanism. Yet, observations of ejecta morphology, spatial
distributions of nucleosynthetic yields and pulsar kicks are
indicative of aspherical (i.e., multi-dimensional) processes bearing
relevance in the explosion dynamics (e.g.,
\cite{wang:08,smith:12a} and references therein). Neutrinos,
on the other hand, are emitted deep inside the core and can provide
crucial thermodynamic, structural, and, to some extent, dynamical
information on what occurs in the core
\cite{bethe:90,lund:10,brandt:11}. The few neutrinos captured from
supernova 1987A \cite{hirata:87,bethe:90} have impressively confirmed
the very basic picture of core collapse outlined in the above.

Gravitational waves (GWs), like neutrinos emitted from dense regions
impenetrable by photons, carry dynamical information about their
source. Since their emission occurs at lowest order by accelerated
quadrupole motions, GWs are direct probes of multi-dimensional
dynamics in the core that may play a crucial role in the supernova
mechanism \cite{ott:09,janka:07}.

Stellar collapse has long been considered a promising source of GWs
for detectors on Earth (see the historical overview in \cite{ott:09})
and much effort has gone into understanding the GW signature of
stellar collapse and the subsequent evolution towards a core-collapse
supernova explosion.  This has led to the identification of a range of
emission processes, including rotating collapse and core bounce,
nonaxisymmetric rotational instabilities, aspherical outflows,
convection/turbulence in the protoneutron star and in the region
immediately behind the shock, instabilities of the standing accretion
shock, pulsations of the protoneutron star, asymmetric emission of
neutrinos, and magnetic stresses (see \cite{ott:09,kotake:12review}
for recent reviews).  The most recent set of simulations
\cite{mueller:12a,muellere:12,kotake:11,dimmelmeier:08,takiwaki:11,ott:09,marek:09}
suggest that GWs from the average core-collapse supernova may be
visible throughout the Milky Way for the second generation of laser
interferometer GW observatories, including Advanced LIGO, Advanced
Virgo, and LCGT \cite{ligo:09,lcgt:10}.  Extreme emission scenarios
may allow detection throughout the local group of galaxies, including
the Andromeda galaxy~\cite{ott:06prl,ott:11a,ott:09,fryernew:11}, but
third-generation detectors such as the Einstein Telescope
\cite{punturo:10} might hope to observe more distant events.

If it was possible to associate an explosion mechanism with particular
multi-dimensional dynamics that leads to a characteristic GW signal,
then the detection or non-detection of such a signal from the next
galactic core-collapse supernova could confirm or rule out this
mechanism. To realize such an GW-observational test of the explosion
mechanism, one must separate the signal from detector noise and
determine its parameters, e.g., by matching, in some way, to signal
predictions from simulations. The most straightforward method for
signal extraction and parameter estimation is matched filtering (e.g.,
\cite{owen:99}), which looks for a match of detector data with
waveform templates from simulations.  Matched filtering requires exact
knowledge of the expected signal, which is possible, e.g., for the
inspiral phase of compact binaries, since the parameter space of
binary systems is limited, all relevant physics is understood (at
least in the black-hole -- black-hole binary case) and numerical
relativity simulations can predict essentially exact waveform
templates (i.e., limited only by numerical error).  However, building
waveform catalogs with exact predictions for matched filtering is
impossible for GWs from core-collapse supernovae.  On the one hand,
there is unknown physics (e.g., the nuclear EOS) and many
unconstrained parameters (e.g., the details of the precollapse
configuration are poorly known) in the stellar collapse problem. This
alone would require extensive parameter studies to build up template
databases covering the poorly constrained parameter space. On the
other hand, all expected GW emission processes in core-collapse
supernovae are influenced or dominated by turbulent flow. Hence, their
GW signals have a stochastic component that is impossible to predict,
even if all initial conditions and physics were known exactly.
Matched filtering is not applicable to such GW bursts. To extract
the GW signal from the next galactic core collapse event and determine
source physics such as the explosion mechanism, an approach to signal
extraction/reconstruction, model selection, and parameter estimation
is needed that can handle the stochastic nature of the expected GW
signals.

The reconstruction of both polarizations of a GW signal requires
coincident observations of two detectors; linearly polarized signals
can be reconstructed from data of just one detector and adding a third
detector overdetermines the problem, permitting the source position on
the sky to be determined \cite{searle:09}. GW signal reconstruction was
pioneered by G\"ursel \& Tinto~\cite{guersel:89} with a maximum
likelihood approach, variants \cite{flanagan:98b,klimenko:05} of which
have been incorporated into search pipelines for GW bursts
\cite{klimenko:08}.

Summerscales \emph{et al.}~\cite{summerscales:08} were the first to study
signal reconstruction and parameter estimation for the GW burst from
rotating core collapse and bounce based on waveforms of
Ott~\emph{et~al.}~\cite{ott:04}. They injected signals into real detector
noise of early LIGO science runs and used a maximum entropy approach
to reconstruct the signal using data from two detectors without any a
priori knowledge of the signal shape. Cross-correlation of the
reconstructed signal with signal predictions of \cite{ott:04} was then
used to determine source parameters. 

Incorporating GW signal information from core collapse simulations
into detection and signal reconstruction approaches was first
considered by Brady \& Ray-Majumder \cite{brady:04}, who realized that
the GW burst from rotating collapse and bounce, while being
unpredictable in detail, has robust features that can be isolated
mathematically. They created an orthonormal vectorspace of waveforms
from \cite{zwerger:97,ott:04} using Gram-Schmidt orthonormalization
and isolated a subspace of essential features most common to all
waveforms. Heng~\cite{heng:09} also considered waveforms from rotating
core collapse and bounce and utilized the more recent waveform catalog
of Dimmelmeier~\emph{et~al.}~\cite{dimmelmeier:08}. He performed
Principal Component Analysis (PCA; e.g., \cite{mardia:79}), which
transforms a correlated, multi-dimensional data set into a set of
orthogonal components by determining the eigenvectors and eigenvalues
of the covariance matrix of the data set. The principal component (PC)
vectors are the eigenvectors ranked according to their corresponding
eigenvalue, with the first PC being the eigenvector with the largest
eigenvalue. Cannon~\emph{et~al.}~\cite{cannon:11} have also utilised
this method for GW signals from compact binary coalescence.

R\"over~\emph{et~al.}~\cite{roever:09} went a step further and
combined PCA with Bayesian inference using the Markov-Chain Monte
Carlo technique for computing marginalization integrals (see, e.g.,
\cite{sivia:06} for a pedagogical introduction to Bayesian
methods). They considered linearly-polarized waveforms from rotating
core collapse and bounce and were able to reconstruct signals from
modeled noise in a single detector and infer key parameters, e.g., the
nuclear EOS used in the simulation that led to a given trial
waveform.

In this paper, we present a proof-of-principle study to demonstrate
that the core-collapse supernova explosion mechanism can be inferred
from the GW signal of a galactic core-collapse supernova observed with
second-generation GW observatories such as Advanced LIGO, Advanced
Virgo, and LCGT. We consider the neutrino mechanism (e.g.,
\cite{janka:07}), the magnetorotational mechanism (e.g.,
\cite{burrows:07b}), and the acoustic mechanism
\cite{burrows:06,burrows:07a}, discuss their essentials, and argue
that they bear distinct GW signatures as first pointed out by
Ott~\cite{ott:09,ott:09b}, which is a prerequisite for our study.  We
follow the approach of R\"over~\emph{et~al.}~\cite{roever:09} and, for
simplicity, restrict ourselves to a single detector, linearly
polarized signals, and a Gaussian noise model at the noise level of
Advanced LIGO in broadband mode. Like R\"over~\emph{et~al.}, we adopt
a Bayesian approach and use PCA, but, for the first time, apply it to
multiple waveform catalogs. We associate each waveform catalog with
one of the three mechanisms and calculate Bayes evidence ratios using the
nested sampling algorithm~\cite{skilling:04,sivia:06} to determine
what mechanism's PCs match best with a given injected signal. We
demonstrate that this approach can identify any of the considered
explosion mechanisms with high confidence for core collapse events
occurring at distances of up to $\sim$$2\,\mathrm{kpc}$.  The
magnetorotational explosion mechanism can even be inferred throughout
the Milky Way ($D \gtrsim 10\,\mathrm{kpc}$).  In addition to studying
the explosion mechanism, we also consider the problem of determining
whether a core collapse event is a rapidly rotating ordinary iron core
collapse or an accretion-induced collapse (AIC) of a massive white
dwarf.  These two processes are governed by the same physics and
differences in their waveforms are subtle~\cite{abdikamalov:10}, but
our approach is still capable of telling them apart.

This article is structured as follows: Section~\ref{sec:snmech}
reviews the set of considered candidate core-collapse supernova
mechanisms, their individual GW signatures, and the GW signal catalogs
that we draw model waveforms from.  In Section~\ref{sec:methods}, we
introduce our method for model selection via PCA and nested
sampling. The results of our study are presented and discussed in
detail in Section~\ref{sec:results}.  We summarize and conclude in
Section~\ref{sec:summary}.


\section{Supernova Mechanisms and their Gravitational Wave
Signatures}
\label{sec:snmech}

In this study, we consider the neutrino mechanism, the
magnetorotational mechanism, and the acoustic mechanism for
core-collapse supernova explosions and describe them and their
characteristic GW signal features in the following sections.
Variations of these mechanisms and alternatives have been discussed
elsewhere (e.g., \cite{ott:12hanse,janka:07}).

\subsection{Neutrino Mechanism}
\label{sec:numech}

The gravitational collapse of the iron core and the subsequent
evolution of the nascent hot puffed-up protoneutron star to a cold
compact neutron star release of order $300\,\mathrm{B}$ (1
$\mathrm{B}$ethe $= 10^{51}\,\mathrm{erg}$) of energy, $\sim$$99\%$ of
which is emitted in the form of neutrinos of all flavors
\cite{bethe:90}. If only a small fraction of the energy released in
neutrinos is re-absorbed behind the stalled shock, leading to net
heating, an explosion could be launched and endowed with the energy to
account for the observed range of asymptotic explosion energies of
$0.1$-$1\,\mathrm{B}$ of garden-variety core-collapse
supernovae~\cite{hamuy:03}. This is the gist of the neutrino mechanism
of core-collapse supernovae, which, in its early form was proposed by
Arnett~\cite{arnett:66} and Colgate \& White \cite{colgate:66}, and in
its modern form by Bethe \& Wilson~\cite{bethewilson:85}.

Despite its appealing simplicity, the neutrino mechanism, in its
purest, spherically-symmetric (1D) form, fails to revive the shock in
all but the lowest-mass massive stars with O-Ne cores
\cite{liebendoerfer:05,rampp:02,thompson:03,kitaura:06}.  There is now
strong evidence from axisymmetric (2D)
\cite{buras:06b,marek:09,mueller:12a,suwa:10,yakunin:10,ott:08,murphy:08}
and first 3D
\cite{fryerwarren:04,nordhaus:10,kuroda:12,muellere:12,kotake:11,iwakami:09,takiwaki:11b,hanke:11}
simulations that the breaking of spherical symmetry is key to the
success of the neutrino mechanism. In 2D and 3D, neutrino-driven
convection in the region of net heating behind the shock, and the
standing-accretion-shock instability (SASI)
\cite{blondin:03,foglizzo:07,foglizzo:06,scheck:08} increase the
efficiency of the neutrino mechanism
\cite{murphy:08,nordhaus:10,kuroda:12,takiwaki:11b}.

\begin{figure}
\centering
\includegraphics[width=1.0\columnwidth]{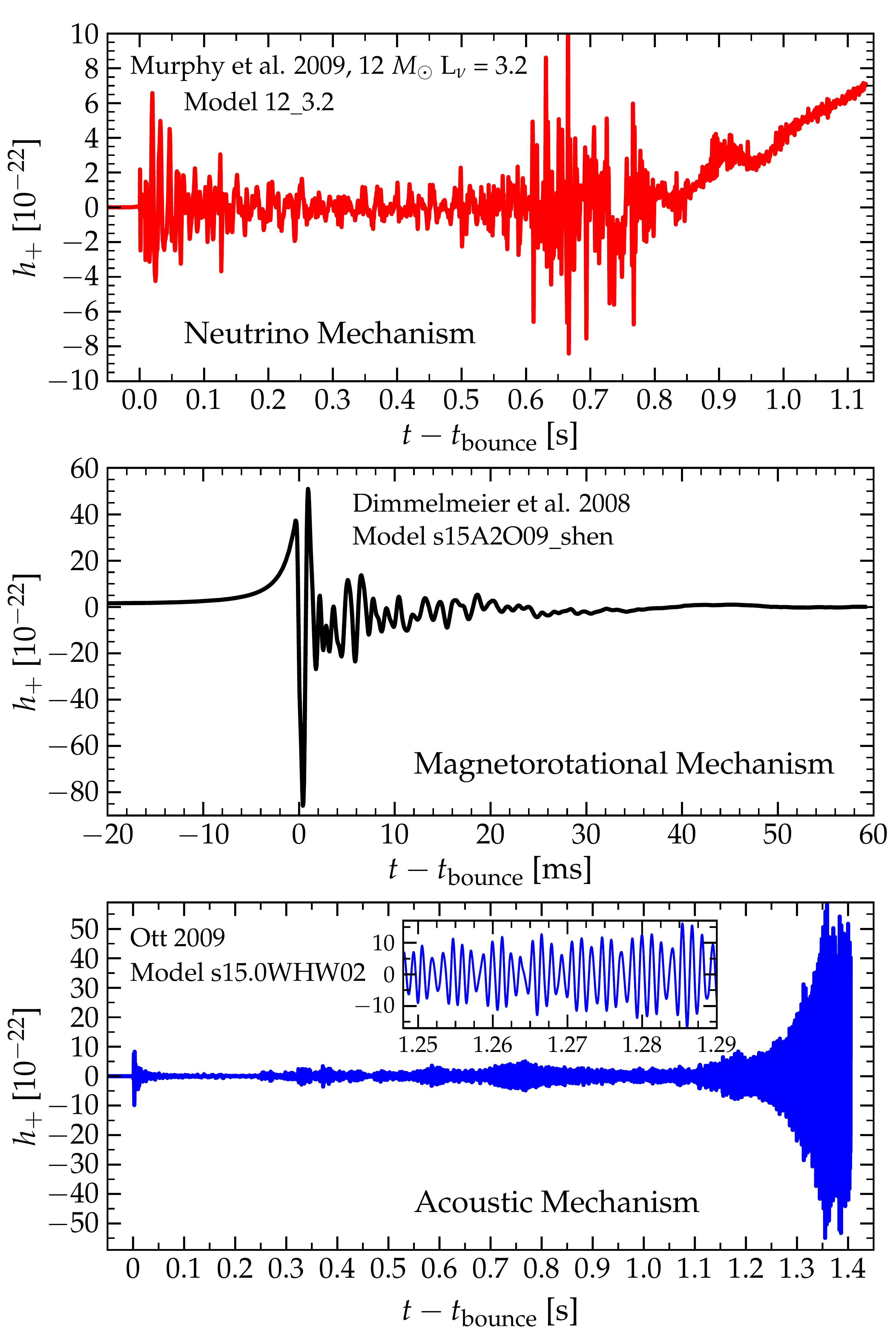}
\caption{Linearly polarized GW signal predictions for a core collapse
  event located at $10\,\mathrm{kpc}$ from matter dynamics
  in axisymmetric simulations that can be associated with the neutrino
  mechanism (top panel, taken from \cite{murphy:09}), the
  magnetorotational mechanism (center panel, taken from
  \cite{dimmelmeier:08}, and the acoustic mechanism (bottom panel,
  taken from \cite{ott:09}). Note the varying ranges of the time
  and strain axes. See text for discussion.}
\label{fig:waveform_examples}
\end{figure}

Leaving rapid rotation aside for a moment, the dominant
multi-dimensional GW-emitting dynamics in neutrino-driven
core-collapse supernovae are convection in the protoneutron star
(e.g., \cite{mueller:04,marek:09b}) and SASI-modulated convection in
the region behind the stalled shock.  GW emission from convection and
SASI has been extensively studied in simulations in 2D
\cite{mueller:04,kotake:07a,marek:09b,murphy:09,yakunin:10} and to
some extent in 3D \cite{fryer:04,muellere:12,kotake:09,kotake:11}.
The top panel in Fig.~\ref{fig:waveform_examples} shows a typical
example waveform drawn from the catalog of Murphy \emph{et
  al.}~\cite{murphy:09}. Right after core bounce, an initial burst of
GWs is emitted by strong, so-called prompt convection \cite{ott:09},
driven by the negative entropy gradient left behind by the stalling
shock. Subsequently, the GW signal settles at lower amplitudes, then
picks up again as the SASI reaches its non-linear phase and
high-velocity accretion downstreams penetrate deep into the region
behind the shock, where they are decelerated, leading to pronounced
spikes in the wave signal~\cite{murphy:09}. The secular rise in the
signal amplitude towards the end of the waveform is due to the onset
of an aspherical explosion \cite{murphy:09,yakunin:10,muellere:12},
but occurs at too low characteristic frequencies to be visible to
Advanced LIGO-class detectors. Not included in the top panel of
Fig.~\ref{fig:waveform_examples} is the contribution to the GW signal
from anisotropic neutrino emission~\cite{epstein:78,bh:96,jm:97},
which can dominate in amplitude, but, like the contribution from
aspherical outflow, occurs on timescales too long to lead to emission
at frequencies detectable by the upcoming generation of ground-based
detectors
\cite{kotake:11,kotake:09,kotake:07a,muellere:12,mueller:04,yakunin:10}.

Overall, the detectable GW signal from a neutrino-driven nonrotating
or slowly rotating core-collapse supernova will have random
polarization, a broadband spectrum from $\sim$$100 - 1000\,
\mathrm{Hz}$ and typical strain amplitudes $|h|$ of order
$10^{-22}\,(D / 10\,\mathrm{kpc})^{-1}$, with individual peaks
reaching $10^{-21} \,(D / 10\,\mathrm{kpc})^{-1}$
\cite{ott:09,muellere:12}.  The typical duration of emission is the
time from core bounce to the launch of the explosion, $0.3 -
1\,\mathrm{s}$, but convection inside the cooling protoneutron star
can continue to emit GWs at lower amplitudes and higher frequencies
for many seconds afterwards \cite{mueller:04,ott:09}.  Typical total
emitted GW energies are in the range $10^{-11}-10^{-9}\,M_\odot\,c^2$
\cite{ott:09,murphy:09,yakunin:10}.

The effects of rotation on the neutrino mechanism and its GW signature
are not yet fully understood (see, e.g.,
\cite{fh:00,suwa:10,marek:09,mueller:12a,ott:08,walder:05,ott:09b}) and it cannot
be excluded that contributions from rotational dynamics may modify the
GW signal of neutrino-driven core-collapse supernovae. However,
results from the systematic rotating core collapse studies of
\cite{dimmelmeier:08,walder:05,burrows:07b,takiwaki:11} suggest that
once rotation rates become sufficiently high to alter the dynamics,
the explosion is actually more likely to occur via the
magnetorotational mechanism discussed in \S\ref{sec:mhdmech}.  This,
however, is under the provision that the magnetorotational instability
(e.g., \cite{bh:91,obergaulinger:09}) works robustly and builds up the
required strong magnetic fields to drive an explosion.

Keeping the above caveats in mind, for the purpose of this study, we
make the assumption that the GW signature of neutrino-driven
core-collapse supernovae is unaffected by rotational effects.

\subsubsection{GW Signal Catalogs}
\label{sec:neutrinocats}

In this study, we use the catalog of
Murphy~\emph{et~al.}~\cite{murphy:09}, which is available for download
from \cite{ottcatalog}. The Murphy~\emph{et~al.}\ catalog (in the
following, we refer to waveforms from this catalog as \cat{Mur}
waveforms) encompasses 16 waveforms that were extracted via the
quadrupole formula (e.g., \cite{thorne:87}) from Newtonian
axisymmetric core collapse simulations that used a parameterized
scheme for electron capture and neutrino heating/cooling and included
only the monopole component of the gravitation potential as described
in \cite{murphy:08,murphy:09}.  The Murphy~et~al. simulations are
nonrotating and the parameter space covered is spanned by progenitor
ZAMS mass (\{$12$, $15$, $20$, and $40$\}~$M_\odot$) and by the
dialed-in total electron and anti-electron neutrino luminosity.

Yakunin~\emph{et~al.}~\cite{yakunin:10} performed self-consistent
axisymmetric Newtonian (with an approximate-GR monopole term of the
gravitational potential~\cite{marek:06}) radiation-hydrodynamics
simulations of neutrino-driven core-collapse supernovae. They provide
three waveforms at \cite{ornlcat}, obtained from
simulations using progenitors of ($12$, $15$, and $25$)~$M_\odot$. We
use the Yakunin waveforms (denoted, in the following, as \cat{Yak}
waveforms) to test the robustness of our supernova mechanism
determination algorithm, which uses the PCs of the \cat{Mur} waveforms.

Since we are limiting ourselves to one detector in this
proof-of-principle study, we are considering only linearly polarized
signals. Gravitational waveforms with $+$ and $\times$
polarizations from 3D simulations of neutrino-driven core-collapse
supernovae \cite{kotake:09,kotake:11,muellere:12} will be considered in
future work.

\subsection{Magnetorotational Mechanism}
\label{sec:mhdmech}

The conservation of angular momentum in core collapse to a
protoneutron star leads to a spin-up by a factor of $\sim$$1000$
\cite{ott:06spin}.  Starting from a precollapse angular velocity
distribution that may be expected to be more or less uniform in the
inner core (e.g., \cite{heger:05}), homologous collapse
preserves the uniform rotation of the inner core while the supersonic
collapse of the outer core leads to strong differential rotation in
the outer protoneutron star and in the region between protoneutron
star and shock \cite{ott:06spin}.

A rapidly spinning precollapse core with a period of order
$1\,\mathrm{s}$ results in a $\mathrm{ms}$-period protoneutron star,
with a rotational energy of order $10\,\mathrm{B}$, which is about ten
times greater than the typical core-collapse supernova explosion
energy.  If only a fraction of this energy was tapped, a strong
explosion could be triggered.

Theory and simulations (e.g.,
\cite{wheeler:00,bisno:70,kotake:04,obergaulinger:06b,shibata:06,burrows:07b,dessart:08a,cerda:08,takiwaki:11})
have shown that magnetorotational processes are efficient at
extracting spin energy and can drive collimated outflows, leading to
energetic bipolar jet-like explosions. Recent work
\cite{shibata:06,burrows:07b,dessart:08a,cerda:08,takiwaki:11}
suggests that magnetic fields of the order of $10^{15}\,\mathrm{G}$
with strong toroidal components are required to yield the necessary
magnetic stresses to drive a strong bipolar explosion. If
$10^{15}\,\mathrm{G}$ fields were to arise from flux compression in
collapse alone, precollapse core fields would have to be of order
$10^{12}\,\mathrm{G}$ \cite{burrows:07b,shibata:06}, which is about 3
orders of magnitude larger than predicted by stellar evolution models
(e.g., \cite{heger:05,woosley:06}). It is more likely that the most
significant amplification occurs after core bounce
via rotational winding of poloidal into toroidal field (a linear
process), the non-linear magnetorotational instability (MRI, which is
not yet fully understood in the core collapse context
\cite{bh:91,obergaulinger:09}). Both processes
operate on the free energy stored in differential rotation, which is
abundant in the outer core.

For the magnetorotational mechanism to work, precollapse spin periods
$\lesssim 4-5\,\mathrm{s}$ appear to be required
\cite{burrows:07b}. Such rapid rotation leads to a strongly
centrifugally-deformed inner core with a large quadrupole moment
($\ell = 2$; due to its oblateness), which rapidly changes during core
bounce, leading to a strong burst of GWs. The GW signal from rotating
collapse and bounce has been studied extensively and the most recent
general-relativistic simulations have shown it to be of rather generic
morphology with a single strong peak at bounce and a subsequent
ringdown as the protoneutron star core settles into its new
equilibrium \cite{dimmelmeier:07,dimmelmeier:08,ott:07prl}. A typical
example GW signal taken from the catalog of
Dimmelmeier~\emph{et~al.}~\cite{dimmelmeier:08,garchingcat} is shown
in the center panel of Fig.~\ref{fig:waveform_examples}. The core
collapse and bounce phase proceeds essentially axisymmetrically even
in very rapidly spinning cores
\cite{ott:07prl,scheidegger:08,scheidegger:10b} and its GW signal is
linearly polarized with vanishing amplitude seen by an observer
located along the symmetry axis and maximum amplitude for an
equatorial observer. Typical emission durations for the linearly
polarized GWs from core bounce are of order $10\,\mathrm{ms}$ and peak
GW amplitudes for rapidly spinning cores that may lead to
magnetorotational explosions are of order $10^{-21} - 10^{-20}$ at
$10\,\mathrm{kpc}$ with most of the energy being emitted around
$500-800\,\mathrm{Hz}$ in cores that reach nuclear density and bounce
due to the stiffening of the nuclear EOS. Cores with initial spin
periods shorter than $\sim$$0.5-1\,\mathrm{s}$ experience a slow
bounce at sub-nuclear densities strongly influenced or dominated by
the centrifugal force. They emit most of the GW energy at frequencies
below $\sim 200\,\mathrm{Hz}$ \cite{dimmelmeier:08,ott:09}.  Typical
emitted GW energies are in the range $10^{-10} -
10^{-8}\,M_\odot\,c^2$. The GW signal from rotating collapse and core
bounce is unlikely to be affected by MHD effects, since the build up
to dynamically relevant field strengths occurs only after bounce
\cite{kotake:04,obergaulinger:06b,takiwaki:11,burrows:07b}.

Due to the strong rotational deformation of the protoneutron star,
neutrinos decouple from the matter at smaller radii and hotter
temperatures in polar regions than near the equator. This leads to the
emission of a larger neutrino flux with a harder neutrino spectrum in
polar regions (e.g., \cite{ott:08}). This globally asymmetric neutrino
emission results in a secularly rising low-frequency GW
signal~\cite{ott:09}. Similar low-frequency contributions will come
from the bipolar outflow characteristic for a magnetorotational
explosion and from magnetic stresses
\cite{kotake:04,obergaulinger:06b,takiwaki:11}.  The low-frequency
waveform components are not shown in the center panel of
Fig.~\ref{fig:waveform_examples} and are not detectable by the
upcoming second-generation earthbound GW observatories.

Also associated with rapid rotation and the magnetorotational
mechanism are rotational instabilities that may lead to
nonaxisymmetric deformations of the protoneutron star whose
``bar-mode'' ($m=2$) components may emit elliptically polarized GWs
for tens to hundreds of milliseconds
\cite{ott:07prl,shibata:05,rmr:98,scheidegger:08,scheidegger:10b}.
However, these instabilities, and in particular their interplay with
magnetic fields and the MRI (see, e.g., \cite{fu:11}), are not yet
fully understood. Since we are considering only linearly polarized
signals and are limited to one detector, we do not include GW
signals from these nonaxisymmetric instabilities in this study.

\subsubsection{GW Signal Catalogs}
\label{sec:mhdcats}

We employ the large (128 waveforms) GW signal catalog of
Dimmelmeier~\emph{et~al.}~\cite{dimmelmeier:08,garchingcat} (\cat{Dim}
in the following), who performed 2D GR simulations of rotating iron
core collapse for ($11.2$, $15$, $20$, and $40$)~$M_\odot$ progenitors
and two different nuclear EOS, varying initial rotation rate and
degree of differential rotation.  They approximated the effects of
electron capture during collapse by parametrizing the electron
fraction $Y_e$ as a function of density, which yields inner core sizes
that are very close to those obtained with full neutrino transport
\cite{liebendoerfer:05fakenu}. The inner core size determines the
amount of mass and angular momentum that can be dynamically relevant
during core bounce and, hence, is a determining factor in the GW
signal \cite{dimmelmeier:07}.  The \cat{Dim} catalog was also used by
the previous parameter estimation work of
R\"over~\emph{et~al.}~\cite{roever:09}.  For testing, we use the three
additional \cat{Dim} waveforms computed for \cite{roever:09} that are
not part of the original \cat{Dim} catalog and were used to test their
algorithm. We label this set of extra waveforms as \cat{DimExtra}.

For studying the robustness of our mechanism-determination approach,
we draw gravitational waveforms of rotating models from the catalog of
Scheidegger~\emph{et~al.}~\cite{scheidegger:10b,scheideggercat},
(\cat{Sch} in the following) who performed 3D Newtonian-MHD rotating
iron core collapse calculations with a spherical approximate-GR
gravitational potential and employed the same EOS and electron capture
treatment as Dimmelmeier~\emph{et~al.}~\cite{dimmelmeier:08}, but used
different progenitor models.

Furthermore, we use the GW signal catalog of
Abdikamalov~\emph{et~al.}~\cite{abdikamalov:10,ottcatalog} (\cat{Abd} in the
following) who used the same numerical code as
Dimmelmeier~\emph{et~al.}~\cite{dimmelmeier:08}, but studied the rapidly
spinning accretion-induced collapse (AIC) of massive white dwarfs to
neutron stars. This process yields a GW signal very similar to
rotating iron core collapse and explosions in AIC may occur also via
the magnetorotational mechanisms \cite{dessart:07aic}. We include this
catalog of 106 waveforms to see if our algorithm can differentiate
between rotating iron core collapse and rotating AIC assuming the \cat{Dim}
and \cat{Abd} catalogs correctly predict the respective GW signals.

\subsection{Acoustic Mechanism}
\label{sec:acousticmech}

The core-collapse supernova evolution in the \emph{acoustic mechanism}
proposed by
Burrows~\emph{et~al.}~\cite{burrows:06,burrows:07a,ott:06prl} is
initially identical to the one expected for the neutrino
mechanism. Neutrino heating, convection and the SASI set the stage,
but no explosion is triggered for $\gtrsim$$500\,\mathrm{ms}$ after
bounce.  At this point, the SASI is in its highly non-linear phase and
modulates high-velocity accretion downflows that impact on the
protoneutron star and excite core pulsations (primarily $\ell =
\{1,2\}$ $g$-modes). Over hundreds of milliseconds, these pulsations
reach large amplitudes and damp via the emission of strong sound
waves. Traveling down the steep density gradient in the region behind
the shock, the sound waves steepen to shocks and dissipate their
energy behind and in the shock. This mechanism is robust in the
simulations by
Burrows~\emph{et~al.}~\cite{burrows:06,burrows:07a,ott:06prl}, but
requires $\gtrsim$$1\,\mathrm{s}$ to develop, thus leads to massive
NSs, and tends to yield explosion energies on the lower side of what
is observed.

The GW signature of the acoustic mechanism is dominated by the strong
emission from the quadrupole components of the protoneutron star core
pulsations that are quasi-periodic (their frequency shifts secularly
along with the changing protoneutron star structure) and become very
strong $\gtrsim$$800-1000\,\mathrm{ms}$ after core bounce
\cite{ott:06prl,ott:09}.  The lower panel of
Fig.~\ref{fig:waveform_examples} depicts a typical example waveform
from Ott~\emph{et~al.}~\cite{ott:09,ott:06prl}, who  studied the
GW signature of the acoustic mechanism based on the simulations of
Burrows~\emph{et~al.}~\cite{burrows:06,burrows:07a}. At early times,
the GW signal is essentially the same as expected for the neutrino
mechanism, but once the protoneutron star core pulsations grow strong,
they are hard to miss. The simulations of
Burrows~\emph{et~al.}~\cite{burrows:06,burrows:07a,ott:06prl} were
axisymmetric and the resulting GW signals are linearly polarized,
though in 3D, one would expect oscillation power also in
non-axisymmetric components. Typical maximum strain amplitudes are of
order $\mathrm{few}\times 10^{-21} - 10^{-20}$ and multiple modes with
frequencies between $\sim 600 - 1000\,\mathrm{Hz}$ contribute to the
emission. Since the pulsations last for many cycles, the emitted GW
energies may be large and are predicted to be of order $10^{-8} -
10^{-7} M_\odot c^2$ and extreme models reach $\mathrm{few}\times
10^{-5} M_\odot c^2$ \cite{ott:06prl,ott:09}.

There are multiple caveats associated with the acoustic mechanism that
must be mentioned. Most importantly, the acoustic mechanism has been
found in simulations of only one group with a single simulation code,
but others have not yet ruled out the possibility of strong
protoneutron star pulsations at late times (e.g., \cite{marek:09}). In
a  non-linear perturbation study, Weinberg \& Quataert
\cite{weinberg:08} found that the protoneutron star pulsation
amplitudes may be limited by a parametric instability involving
high-order modes that damp efficiently via neutrino emission and are
not presently resolved in numerical simulations. This would limit the
protoneutron star pulsations to dynamically insignificant
amplitudes. Moreover, the simulations of Burrows~\emph{et~al.}  were
axisymmetric and nonrotating or only very slowly rotating. It is not
clear to what amplitudes individual protoneutron star pulsation modes
would grow in 3D. Rapid rotation, due to its stabilizing effect on
convection and SASI \cite{ott:08,ott:09b}, may likely inhibit the
growth of pulsations. Both 3D and rotational effects remain to be
explored.

\subsubsection{GW Signal Catalogs}
\label{sec:acousticcats}
We employ the set of 7 waveforms from the models of \cite{burrows:07a}
analyzed by Ott \cite{ott:09} and available at \cite{ottcatalog}.  We
refer to this set as the \cat{Ott} catalog in the following and use
them to compute PCs for the acoustic mechanism's GW signature. All
waveforms were computed on the basis of the
Burrows~\emph{et~al.}~\cite{burrows:06,burrows:07a,ott:06prl} simulations and
differ only in the employed progenitor model, covering a range in ZAMS
mass from $11.2$ to $25\,M_\odot$.

Three additional waveforms of an earlier study of Ott~\emph{et al.}
\cite{ott:06prl} are available \cite{ottcatalog}. We label this small
set \cat{OttExtra} and use them for testing our method's capability of
correctly identifying them as coming from stars exploding via the
acoustic mechanism.


\section{Data Analysis}
\label{sec:methods}

\subsection{Strategy}

The three example gravitational waveforms shown in
Fig.~\ref{fig:waveform_examples} that are associated with the three
supernova mechanisms are clearly different. Provided the assumptions
made in associating these signals with the various mechanisms are
correct, a GW signal detected from a core-collapse supernova should,
in principle, allow to determine the explosion mechanism. To do so in
practice, two problems most be overcome: (\emph{i}) The exact waveform
of an incident signal is impossible to predict in advance.
(\emph{ii}) Real GW detectors are noisy instruments (see, e.g.,
\cite{pitkin:11} for a discussion of detectors and noise sources) and
any GW signal will be contaminated by detector noise.
In other words, it is necessary to develop a data analysis algorithm
that is capable of distinguishing between underlying physical models
(e.g., supernova mechanisms) on the basis of a noisy signal
whose detailed shape cannot be predicted exactly.

\begin{figure}
\centering
\includegraphics[width=8cm]{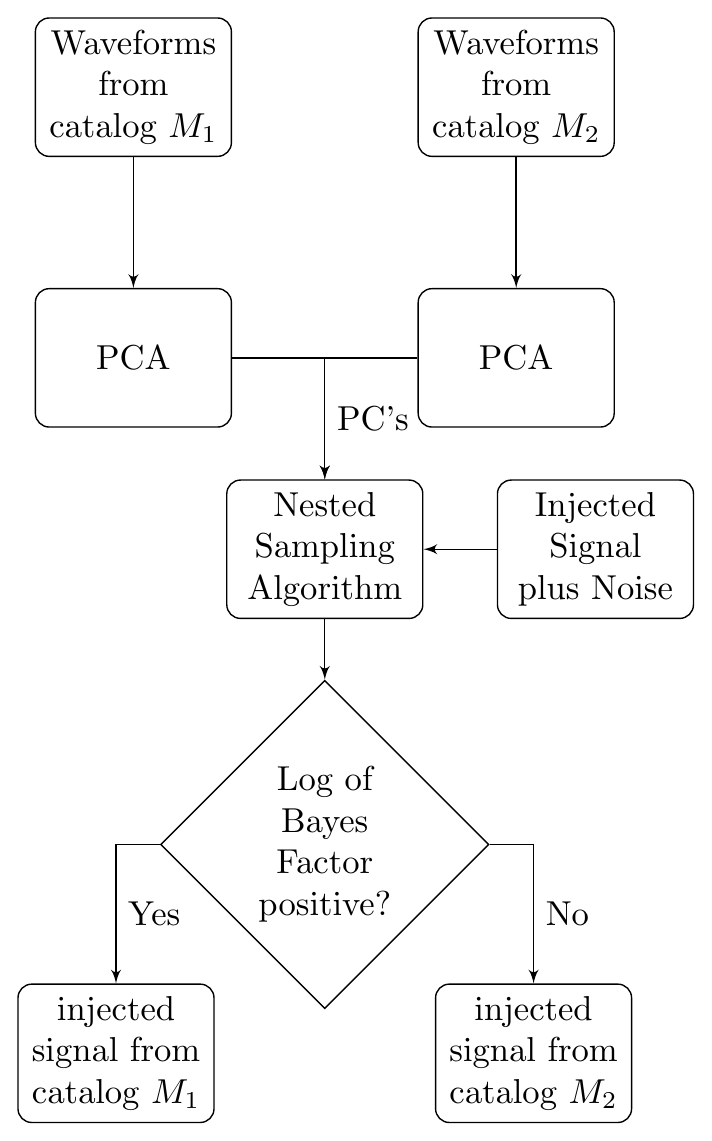}
\caption{Block diagram of the Supernova Model Evidence Extractor
  (SMEE). A desired core-collapse supernova gravitational wave signal
  is injected into noise, and the algorithm compares it to the
  principal components (PCs) of a given waveform catalog representing
  a particular model.  The PCs are constructed via singular value
  decomposition (SVD).  The sign of the log Bayes factor between two
  PC sets indicates which model is favored by the data.}
\label{fig:flowchart}
\end{figure}

In the following subsections, we describe the components of a Bayesian
data analysis algorithm which classifies detected GW signals from
core-collapse supernovae as belonging to one of a set of signal
catalogs, representing, e.g., different explosion mechanisms.  A block
diagram of the analysis algorithm, which we call the Supernova Model
Evidence Extractor (SMEE), is shown in
Fig.~\ref{fig:flowchart}. SMEE is implemented in {\tt
  MATLAB}\footnote{The MathWorks Inc., Natick, MA 01760, USA. {\tt
    http://www.mathworks.com/products/matlab/}.}.

In a first step, SMEE performs Principal Component Analysis (PCA) via
singular value decomposition (SVD) on the waveforms in each catalog to
create sets of orthogonal basis vectors, the Principal Components
(PCs), which are ordered according to their prevalence in their
catalog. In other words, the first PC represents the most common
feature of all signals in the catalog, the second PC represents the
second most common feature, and so on. Using a complete set of PCs,
each waveform can be reconstructed as a linear combination of PCs for
the corresponding catalog, allowing each waveform to be simply
parameterized by the PC coefficients in the linear
combination. However, since PCs are expected to span the parameter
space defined by each catalog of waveforms efficiently, catalog
waveforms may be reconstructed with good accuracy already with a set
of PCs that is significantly smaller than the number of waveforms in
the catalog.  Moreover, non-catalog waveforms (i.e., real signals) may
be identified as belonging to the same class of signals as catalog
waveforms if they can be approximately matched with the first few PCs
of a catalog.

SMEE then uses Bayesian model selection and computes the logarithm of
the Bayes factor to distinguish between GW signal classes.  The Bayes
factor is the ratio of the evidences for two competing hypotheses and,
for the purpose of our analysis, we weigh the evidence that the
observed data supports the presence of a GW signal consistent with
signals from one of any two competing catalogs. This requires summing
up the likelihood function times the prior across all possible signal
parameters (in our case, values of PC coefficients) to determine the
evidence (also called the marginal likelihood) for two different
signal models to be tested. SMEE accomplishes this efficiently via the
Nested Sampling algorithm \cite{skilling:04,sivia:06}.

\subsection{Bayesian Model Selection}
\label{sec:bayessection}

In our analysis, we employ Bayesian Model Selection, similar to that
described in~\cite{delpozzo:11}. Specifically, we use the Bayes factor
to compare the probabilities of two competing models. In general terms,
the Bayes factor $B_{ij}$ can be written as the ratio of the
evidences $p(D|M)$,

\begin{equation}
B_{ij}=\frac{p(D|M_i)}{p(D|M_j)}\,\,,
\end{equation}

where $M_i$ and $M_j$ are two competing models tested using the data
$D$. The evidence for each model is obtained by integrating the
product of its likelihood function and prior across all model
parameter values $\theta$, such that,

\begin{equation} 
\label{eq:evi} 
p(D|M)=\int\limits_\theta p(\theta|M)p(D|\theta,M) \, d\theta\,\,.
\end{equation}

The evidence will be greater for a model that is supported by the
data. Therefore, the Bayes factor indicates which of the two competing
models is preferred by the data.  It is often more convenient to
compare models by using the natural logarithm of the Bayes factor,

\begin{equation} 
\label{eq:bayes} 
\log B_{ij} = \log{p(D|M_i)} - \log{p(D|M_j)}\,\,. 
\end{equation}

In this case, $\log{B_{ij}} > 0$ means $M_i$ is the preferred model
whereas $\log{B_{ij}} < 0$ will point to $M_j$ being favored.

\subsection{Nested Sampling Algorithm}

For evaluating $\log{B_{ij}}$, we first need to calculate the
evidences $p(D|M_i)$ and $p(D|M_j)$ for the two models $M_i$ and
$M_j$.  From Eq.~\ref{eq:evi} we see that the evidence is the sum of
the likelihood times the prior determined for all possible parameter
values of the desired model.  An exhaustive, brute-force approach to
computing the evidence by calculating the likelihood values for every
choice of parameter values is computationally prohibitive.  It is also
an inefficient way of determining the evidence since the likelihood
values will be most significant, and therefore contribute most to the
evidence, for a small subset of parameter values which constructs a
waveform that closely resembles the data.  For most other combinations
of the model's parameters, the likelihood will be insignificant and
not contribute to the evidence.

Therefore, we choose to follow the approach of Veitch
\emph{et al.}~\cite{veitch:10} and employ Nested
Sampling~\cite{skilling:04,sivia:06} to efficiently calculate the
evidence integral.  The Nested Sampling algorithm determines the
evidence integrals by calculating the likelihood for a selected sample
of parameter values for the desired model.  Initially, the model's
parameter values are randomly selected before the algorithm
iteratively converges on the set of parameter values that produce the
most significant likelihood values.  It is similar to the Markov-Chain
Monte Carlo approach (e.g., \cite{sivia:06}) except that the primary
output of the Nested Sampling algorithm is the evidence, which can be
immediately obtained by summation whereas the posterior distribution
is only found as a by-product. To find the evidence, a set of ``live
points'' are found through creating a stochastic sampling of the prior
distribution to generate a set of $N$ samples which are denoted as
$\theta_{i}$, where $i=1 \ldots N$. The evidence integral
(Eq.~\ref{eq:evi}) is then written as

\begin{align} 
p(D|M)&=\int\limits_\theta p(\theta|M)p(D|\theta,M) \, d\theta\,, \nonumber \\
               &\approx \sum_{i=1}^{N} p(D|\theta_{i},M)w_{i}\,, \nonumber \\ 
	       &\approx \sum_{i=1}^{N} L_{i}w_{i}\,,
\end{align}
where the weight,
\begin{equation} 
w_{i}= p(\theta_{i}|M) d\theta_{i}\,,
 \end{equation}
 
 is the fraction of the prior distribution represented by the $i$-th
 sample and $L_{i}$ is its likelihood.  It is this weighted likelihood
 that is calculated by the Nested Sampling algorithm and subsequently
 used to obtained the evidence. More details on the Nested Sampling
 algorithm can be found in \cite{veitch:10,skilling:04}.


\subsection{Principal Component Analysis via 
Singular Value Decomposition} \label{sec:svd}

Each core-collapse supernova waveform catalog consists of a number of
GW signals obtained for different initial conditions and simulation
parameters (e.g., progenitor star mass, EOS, rotational configuration
etc.). While individual waveforms of one catalog are different in
detail, they generally exhibit strong common general features. This
can be exploited by principal component analysis (PCA)
\cite{mardia:79}, which isolates the most common features of waveforms
in linearly independent principal components (PCs) ordered by their
relevance. The first few PCs may already be sufficient to efficiently
span their entire catalog, as was shown in \cite{heng:09} and
\cite{roever:09} for the \cat{Dim} catalog (see Sect.~\ref{sec:mhdcats}).

The PCs are obtained via singular value decomposition (SVD) (e.g.,
\cite{mardia:79}) of time-domain waveforms from each catalog.  
To perform SVD on a catalog with $m$ waveforms, a
matrix {\bf A} is created such that each of its columns corresponds
to a waveform of uniform length $n$ from the catalog. 

The $n \times m$ waveform matrix {\bf A} is factored so that

\begin{equation}
\label{eq:svd}
{\bf A} = {\bf U \Sigma V}^T,
\end{equation}

where {\bf U} is an $n \times n$ matrix whose columns correspond to
the eigenvectors of ${\bf AA}^T$. Similarly, the columns of the $m
\times m$ matrix {\bf V} correspond to the eigenvectors of ${\bf
  A}^T{\bf A}$ and ${\bf \Sigma}$ is an $n \times m$ diagonal matrix
whose elements correspond to the square root of the corresponding
eigenvalues.

Since ${\bf AA}^T$ is the covariance matrix of {\bf A}, the
eigenvectors in {\bf U} are effectively an orthonormal basis which
span the $m$-dimensional parameter space defined by the catalog of
waveforms used to construct {\bf A}. Note that, in practice, $n \gg m$
and it is impractical to determine the eigenvectors in {\bf U}
directly. Instead, the smaller {\bf V} and its corresponding
eigenvalues in ${\bf \Sigma}$ are first determined which are
subsequently used to derive {\bf U}.

The orthonormal eigenvectors of {\bf U} are the PCs and are
ranked by their corresponding eigenvalues. The PC with the largest
corresponding eigenvalue is referred to as the first PC and consists
of the most significant common features of all waveforms in the
catalog. It follows that the PC with the second largest corresponding
eigenvalue is the second PC and consists of the second most
significant common features and so on. 

The waveforms in {\bf A} can be reconstructed by taking a linear
combination of PCs, 

\begin{equation}
\label{eq:beta}
h_i \approx \sum_{j=1}^{k}{U_j}{\beta_j}\,\,,
\end{equation}

where $h_i$ is the desired waveform from the catalog, $U_j$ is the
$j$th PC from the {\bf U} matrix and $\beta_j$ is the corresponding PC
coefficient, which can be obtained by projection of $h_i$ onto $U_j$. The
sum of $k$ PCs produces an approximation of the desired waveform since
$k \le m$.

\subsection{Signal and Noise Models}
\label{sec:signal_and_noise_models}

For the analysis described here, two types of models are considered.
The signal model $M_s$ tests the presence of a signal waveform
$h(\beta)$ in the data. Here, PCA is performed for each catalog (using
SVD; see section~\ref{sec:svd}) and each waveform is
parameterized by its PC coefficients ($\beta$).  The Gaussian
likelihood function for the signal model is

\begin{equation}
\label{eq:like}
p(D|\beta,M_s)= \prod_{i=1}^{N}  \frac{1}{\sigma_i\sqrt{2\pi}} \exp \left [ -\frac{(D_i-h_i(\beta))^2}{2 \sigma^2_i} \right ],
\end{equation}

where $\sigma_i$ is the standard deviation of the noise, $h_i(\beta)$
is the desired waveform reconstructed from the PCs and $N$ is the
length of the data with a corresponding index $i$.  The evidence for
the signal model is determined by performing the integral in
Eq.~\ref{eq:evi} numerically across all chosen values of $\beta$.

On the other hand, the noise model $M_n$ tests the data's consistency
with Gaussian noise. The likelihood function for the noise model is
the same as that in Eq.~\ref{eq:like}, but with $h(\beta)=0$. From
this, it is straightforward to perform the integration in
Eq.~\ref{eq:evi} and obtain an analytic form for the noise evidence
function,

\begin{equation}
\label{eq:noise}
p(D|M_n) = \prod_{i=1}^{N}  \frac{1}{\sigma_i\sqrt{2\pi}} \exp^{-\frac{D^2_i}{2\sigma^2_i}}.
\end{equation}

In both Eq.~\ref{eq:like} and \ref{eq:noise}, the standard deviation
of the noise is a function of each sample in the data, because the
simulated noise is designed to correspond to the expected sensitivity
of Advanced LIGO, which varies as a function of frequency (see
Sec.~\ref{sec:noisegen}). To handle the frequency-dependent noise, the
signal and noise evidences are calculated in the frequency domain with
$D_i$, $h_i(\beta)$ and $\sigma_i$ corresponding to data,
reconstructed waveform, and noise in the $i$-th frequency bin
respectively. In particular, each reconstructed waveform is obtained
by taking a linear combination of the Fourier transforms of its
corresponding PCs.

The natural logarithm of the Bayes factor used to compare the
signal model to the noise model is then simply 
\begin{align}
\label{eq:sigvsnoise} 
\log{B_{SN}} &= \log[p(D|M_s)] - \log[p(D|M_n)] .
 \end{align}

\begin{figure*}[t]
\centering
\includegraphics[width=1.0\textwidth]{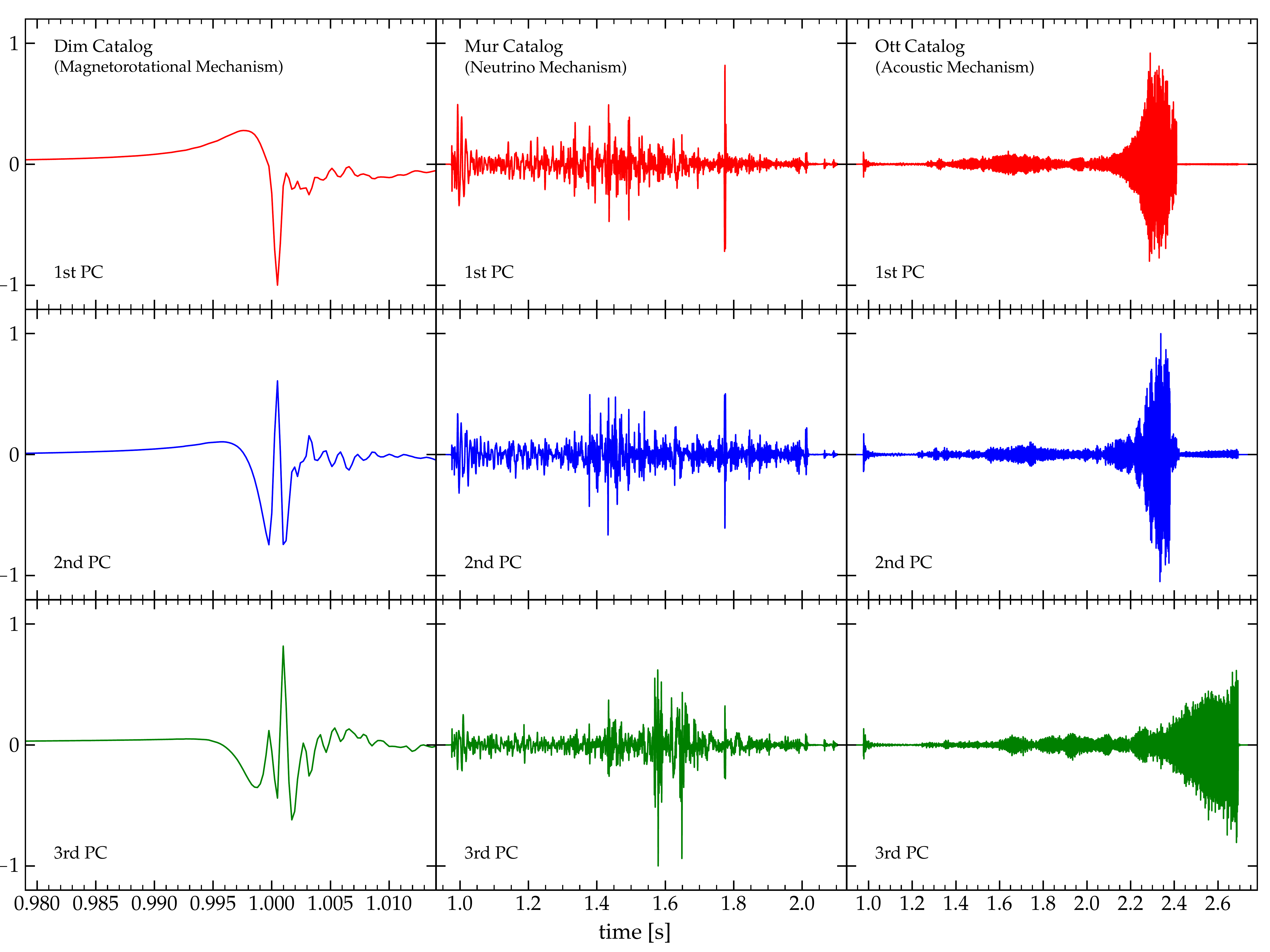}
\caption{The first three Principal Components (PCs) of the waveforms
  from the Dimmelmeier (\cat{Dim}) \cite{dimmelmeier:08,garchingcat},
  Murphy (\cat{Mur}) \cite{murphy:09,ottcatalog}, and \cat{Ott}
  \cite{ott:09,ottcatalog} catalogs, which we take to be
  representative of the magnetorotational, neutrino and
  acoustic mechanisms, respectively (see
  Secs.~\ref{sec:numech},~\ref{sec:mhdmech}
  and~\ref{sec:acousticmech}).  In creating the PCs, the waveforms of
  each catalog are placed in a systematic way in a 3 s interval and
  padded left and right by zeros. The vertical axis is a dimensionless
  scale which represents the amplitude,  which we have
  normalized by the maximum amplitude over all 3 PCs of each catalog
  shown here.}
\label{fig:AllPCs}
\end{figure*}

\subsection{Generation of Simulated Noise}
\label{sec:noisegen}

We generate Gaussian colored noise, assuming a
single Advanced LIGO detector in the proposed broadband configuration
(the so-called ``zero detuning, high-power'' mode). We employ the data
file {\tt ZERO\_DET\_high\_P.txt} provided by \cite{LIGO-sens-2010},
which contains $\sqrt{S(f)}$, the square root of the one-sided
detector noise power spectral density in units of
$(\mathrm{Hz})^{-{1/2}}$. An open-source implementation for {\tt MATLAB} of
what we describe in the following can be found in
\cite{matappswebnoise}.

The real discrete time-domain noise $n(t_j)$, where $t_j$ denotes the
$j$-th discrete time interval of size $\Delta t$, is obtained by
inverse discrete Fourier transform from the complex frequency-domain
noise $\tilde{n}(f_k)$, where $f_k$ denotes the $k$-th discrete
frequency interval of size $\Delta f = 1 / (N_t \Delta t)$, where $N_t$ is
the number of intervals in the time domain. Since the time-domain
noise is real and has zero mean, the frequency-domain noise must obey
\begin{eqnarray}
\label{eq:ff1}
\tilde{n}(-f) &=& \tilde{n}^*(f)\,\,,\\\
\label{eq:ff2}
\tilde{n}(f=0) &=& 
\tilde{n}(f=f_\mathrm{Nyq}) = \tilde{n}(f=-f_\mathrm{Nyq}) = 0. 
\end{eqnarray}
Here, $f_\mathrm{Nyq} = 1 / (2\Delta t)$ is the Nyquist frequency. If
$N_t$ is the even number of equally spaced bins in time of width
$\Delta t$, then $N_f = N_t/2 - 1$ is the number of independent
frequency bins $f_k$ in the frequency domain of width $\Delta f$. The
frequency variable $f_k$ assumes values from
$-f_\mathrm{Nyq}$ to $f_\mathrm{Nyq}$.

$|\tilde{n}(f_k)| = \sqrt{\tilde{n}(f_k) \tilde{n}^*(f_k)}$ is a two-sided
amplitude spectral density. We generate $\tilde{n}(f_k)$ by sampling the
standard normal distribution (zero mean, variance one) weighted by the
noise transfer function $T(f_k) = \sqrt{S(f_k)} / \sqrt{2}$. The real and
imaginary parts of $\tilde{n}(f_k)$ are then given for each $f_k \in
(0,f_\mathrm{Nyq})$ by
\begin{eqnarray}
\mathbb{R}(\tilde{n}(f_k)) &=& \frac{1}{\sqrt{2}}\,\, T(f)\,\,{\tt RANDN}\,\,,\\
\mathbb{I}(\tilde{n}(f_k)) &=& \frac{1}{\sqrt{2}}\,\, T(f)\,\,{\tt RANDN}\,\,,
\end{eqnarray}
where ${\tt RANDN}$ is a random number sampled from the standard normal
distribution. The remaining $\tilde{n}(f_k)$ are then obtained via
Eqs.~\ref{eq:ff1} and \ref{eq:ff2}.

The inverse discrete Fourier transform of $\tilde{n}(f_k)$ to time
domain noise $n(t_j)$ will preserve its Gaussian character in the time
domain in the limit of small sampling interval
\cite{roever:09,veitch:10}.  Specifically, we use the following
definition of the discrete Fourier transform (DFT) for transforming
noise from the frequency to the time domain when needed:
\begin{align}
\label{eq:dft}
\tilde{n}(f_k) &= \sum_{j=0}^{N-1} n(t_j) \mathrm{exp}(-2\pi \mathrm{i}\, j k / N)\,\,, \\
n(t_j) &= \frac{1}{N} \sum_{k=0}^{N-1} \tilde{n}(f_k) \mathrm{exp}(2 \pi \mathrm{i}\, j k / N)\,\,. 
\end{align}

For convenience, we define the matched filter signal-to-noise ratio
(SNR) of a GW signal $h$ as
\begin{align}
\mathrm{SNR}^2 &= 4 \int_{0}^{\infty}
\frac{\abs{\tilde{h}(f)}^2}{S(f)} df \\ 
&= 4\, \Delta t^2 \Delta f
\sum_{k=1}^{N_f} \frac{\abs{\tilde{h}(f_{k})}^2}{S(f_{k})}\,\,,
\label{eq:SNR}
\end{align}
where $S(f)$ is the one-sided noise power spectral density. The factor
$\Delta t^2$ is applied to correct the dimensions of $\tilde{h}(f)$,
which we obtain via the DFT defined by Eq.~\ref{eq:dft}.

\subsection{Application in SMEE}

\subsubsection{GW Signal Preparation and PCA}

Before carrying out PCA and injecting signals into noise, all
waveforms are buffered with zeros to be of length $n$, which we choose
to correspond to $3\,\mathrm{s}$ at a sampling rate of
$4096\,\mathrm{Hz}$, allowing us to comfortably accommodate the
longest available core-collapse supernova GW signals. The Advanced
LIGO sampling rate is $16\,\mathrm{kHz}$.  The reduced sampling rate
we choose saves computation time and is sufficient to capture the
frequency content of the core-collapse supernova waveforms considered
here, which have most of their power at $\sim$$50 -
1000\,\mathrm{Hz}$.

We align waveforms from the \cat{Dim} catalog at their maximum (the
spike at core bounce). Waveforms from the \cat{Mur} and \cat{Ott}
catalogs are aligned so that the onsets of emission coincide. All
waveforms are shifted so that they are aligned to the $4000$-th point
in the SMEE input data file, corresponding to about the
$1\,\mathrm{s}$ mark in the $3\,\mathrm{s}$ interval, to leave ample
space left and right of the waveform.

In Fig.~\ref{fig:AllPCs}, we present the first three PCs computed for
the \cat{Dim} (magnetorotational mechanism; left panel), \cat{Mur} (neutrino
mechanism; center panel), and \cat{Ott} (acoustic mechanism; right panel)
catalogs. Before generating PCs for the \cat{Mur} catalog we filter out the
secular low-frequency drifts present in the \cat{Mur} waveforms (see
Fig.~\ref{fig:waveform_examples}) by high-passing the signal above
$30\,\mathrm{Hz}$. Since the low-frequency components are hidden in
detector noise even when the source is nearby, dropping them improves
the efficiency of our subsequent Bayesian analysis and signal
reconstruction. We apply the same high-passing to trial waveforms for
the neutrino mechanism before injecting them into noise.

\subsubsection{Signal Injection and Model Selection}

We inject trial GW signals into simulated Gaussian Advanced LIGO noise
and use SMEE to determine which signal model (e.g., what core-collapse
supernova explosion mechanism) a given injected signal belongs to
via the evaluation of the logarithmic Bayes factors $\log B_{SN}$
(Eq.~\ref{eq:sigvsnoise}) for an injected signal for each signal model
$S$ and the noise model $N$. Comparing two signal models $i$ and $j$
is then accomplished by computing $\log B_{ij} = \log B_{iN} - \log
B_{jN}$.

SMEE's model selection operates in the frequency domain. Trial GW
signal and the PCs belonging to the signal model under consideration
are transformed into the frequency domain via DFT and the trial GW
signal is added to the complex frequency-domain noise, retaining phase
information.  The Nested Sampling algorithm is then invoked to
marginalize the PC coefficients $\beta_k$. The prior for each
coefficient is flat and uniform. The prior range for each $\beta_k$ is
determined by first reprojecting all waveforms of a given catalog back
onto the PCs to compute $\hat{\beta}_{kl}$ for each PC $k$ and
waveform $l$ of the catalog. The range of expected possible values of
$\beta_k$ is then found by taking the minimum and maximum of
$\hat{\beta}_{kl}$ over all $l$ and adjusting these numbers by $10\%$
down and up to add a margin of error to account for uncertainty due to
the noise, motivated by the findings of \cite{roever:09}.

Keeping the noise model fixed, the results of SMEE's computations will
depend on the SNR of the signal, i.e.\ the distance to the core
collapse event, and on the amount of information we can provide to
SMEE about expected signals in the form of PCs.

The maximum number of PCs at SMEE's disposal is limited by the number
of waveforms used to determine the set of PCs.  While each catalog
used in this study has a different number of waveforms, we choose to
simplify our analysis by using the same number of PCs for all
catalogs. Hence, the maximum number of PCs we use here is $7$ and is
set by the number of waveform in the \cat{Ott} catalog (see
\S\ref{sec:acousticcats}).  Using $7$ PCs gives SMEE complete
information about signals belonging to the \cat{Ott} catalog and
significant, but incomplete information about waveforms from the
\cat{Dim}, \cat{Mur}, and \cat{Abd} catalogs. We also carry out SMEE
runs with less than $7$ PCs to study the dependence on the number of
PCs employed.  Using only a small subset of a catalog's PCs limits
SMEE's ability to precisely reconstruct injected catalog waveforms,
but it represents the real-life situation that the a priori
information about a detected signal is severely limited. Our goal here
is not to ideally reconstruct signals but to show that determining the
underlying physical model of an observed signal is possible with
limited advance knowledge.


\section{Results}
\label{sec:results}

\subsection{Response to Gaussian Noise}
\label{sec:purenoise}

\begin{figure}[t]
\centering
\includegraphics[width=0.5\textwidth]{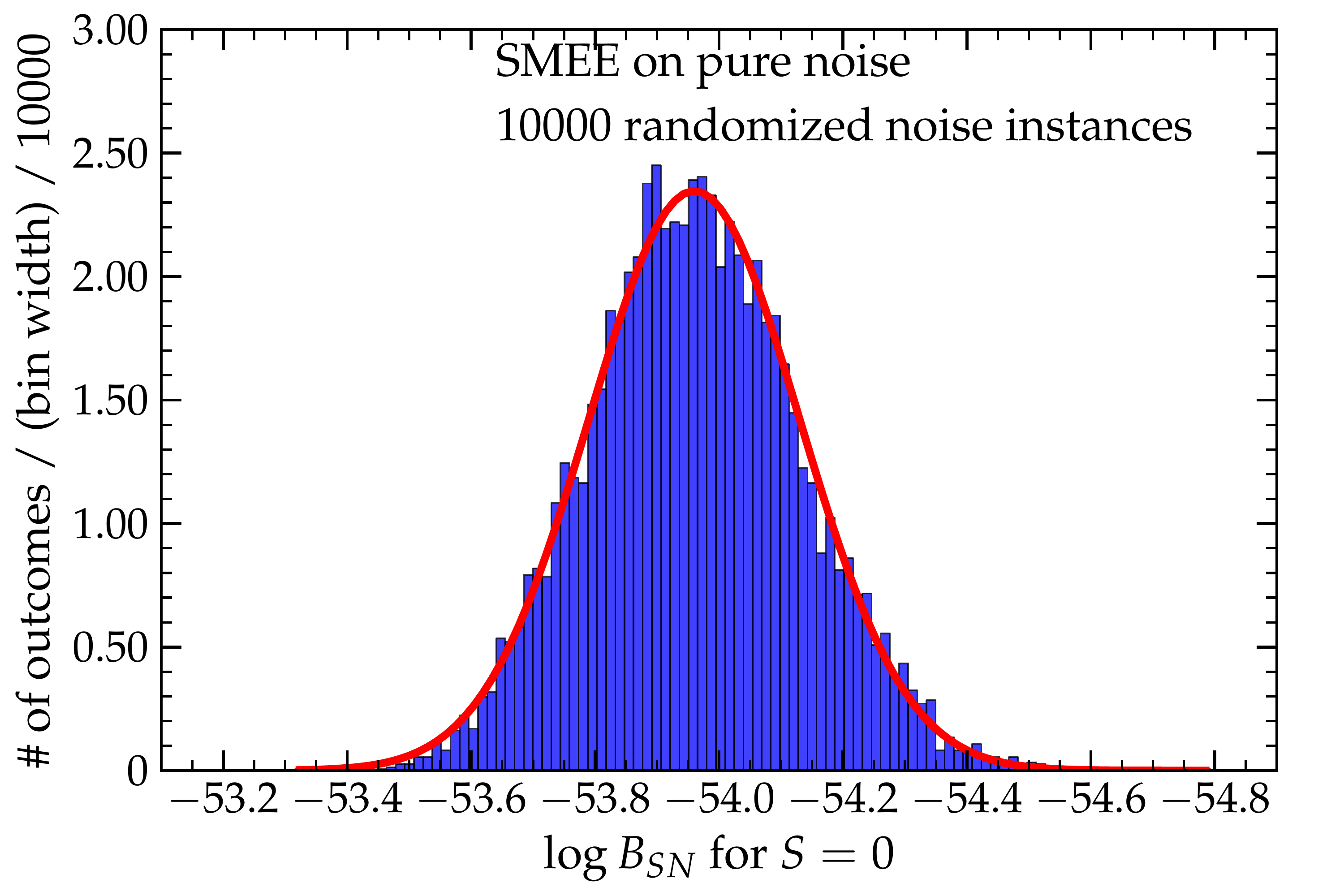}
\caption{Results from running SMEE with 7 PCs of the \cat{Dim} catalog
  and without an injected signal on 10,000 randomized instances of
  Gaussian Advanced LIGO noise, generated as described in
  Sec.~\ref{sec:noisegen}. A signal consistent with noise is most
  likely to have a logarithmic Bayes factor of $\sim$$-54.0$. The red line
  plots a Gaussian distribution with a mean of $-53.96$ and a standard
  deviation $\sigma = 0.17$.}
\label{fig:noise}
\end{figure}

For interpreting the results of SMEE's Bayesian model selection on the
basis of Eq.~\ref{eq:bayes}, it is necessary to quantify and
understand SMEE's response to pure Gaussian detector noise without a
signal being present.  To this end, we run SMEE on 10,000 randomized
instances of Advanced LIGO detector noise (generated as described in
\S\ref{sec:noisegen}) without injecting signals and compute $\log
B_{SN}$ (Eq.~\ref{eq:sigvsnoise}) in the absence of a signal for each
signal model $S$. The results, shown in Fig.~\ref{fig:noise}, follow a
Gaussian distribution with a mean corresponding to the expected value
$-\sum_{i=1}^N \frac{h_i(\beta)^2}{2\sigma_i^2}$, where $N$ is the
number of PCs employed (see \S\ref{sec:signal_and_noise_models}).
The average logarithmic Bayes factors obtained for 10,000
  instances of noise indicate that noise, or any signal fully
  consistent with noise, is most likely to have a logarithmic Bayes
  factor of $-54.0$ when SMEE is run with 7 PCs of the \cat{Dim}
  catalog. For the \cat{Ott}, \cat{Mur}, and \cat{Abd} catalogs, the
  expected logarithmic Bayes factors for pure Gaussian noise and 7 PCs
  are $-52.1$, $-52.3$, and $-53.0$, respectively.  The observed
  expectation value are very comparable to those calculated for the
  \cat{Dim} ($-53.9$), \cat{Ott} ($-52.2$), \cat{Mur} ($-52.3$), and
  \cat{Abd} ($-52.9$) catalogs, respectively, verifying that SMEE is
  operating as expected. We have repeated this experiment for the case
when only 3 PCs are used and also in this case find that SMEE closely
reproduces the predicted expectation values, which are near $-26$ in the
3-PC case.

Since the logarithmic Bayes factors follow a Gaussian distribution, we
can set a threshold using the standard deviations as an indicator for
the expected false alarm rate. Ideally, for the \cat{Dim} catalog, a
$1\%$ false alarm rate would correspond to a threshold that is
$\sim$$2.6$ times the standard deviation, corresponding to
$\sim$$0.44$ above the mean. However, we note that the expected
logarithmic Bayes factor value varies between different catalogs and,
for a fixed false alarm rate, we would require a different threshold
for each catalog. This variation can be address by re-normalising all
Bayes factors so that they are the same for all catalogs when there is
only noise. But, since the focus of our work here is to distinguish
between different waveforms and not to perform a study on the
detection efficiency of GW signals, we choose to take the more
conservative approach of simply setting a higher threshold.
Therefore, we conservatively choose to identify a signal
  as being distinct from noise if its $\log B_{SN}$ is greater than
  $-47$ (in cases in which we use 3 PCs, this number is $-21$). When
  comparing two signal models $M_i$ and $M_j$, we conservatively
  identify model $M_i$ as favored if $\log B_{ij} \ge 5$ (and vice
  versa).

\begin{figure}[t]
\centering
\includegraphics[width=\columnwidth]{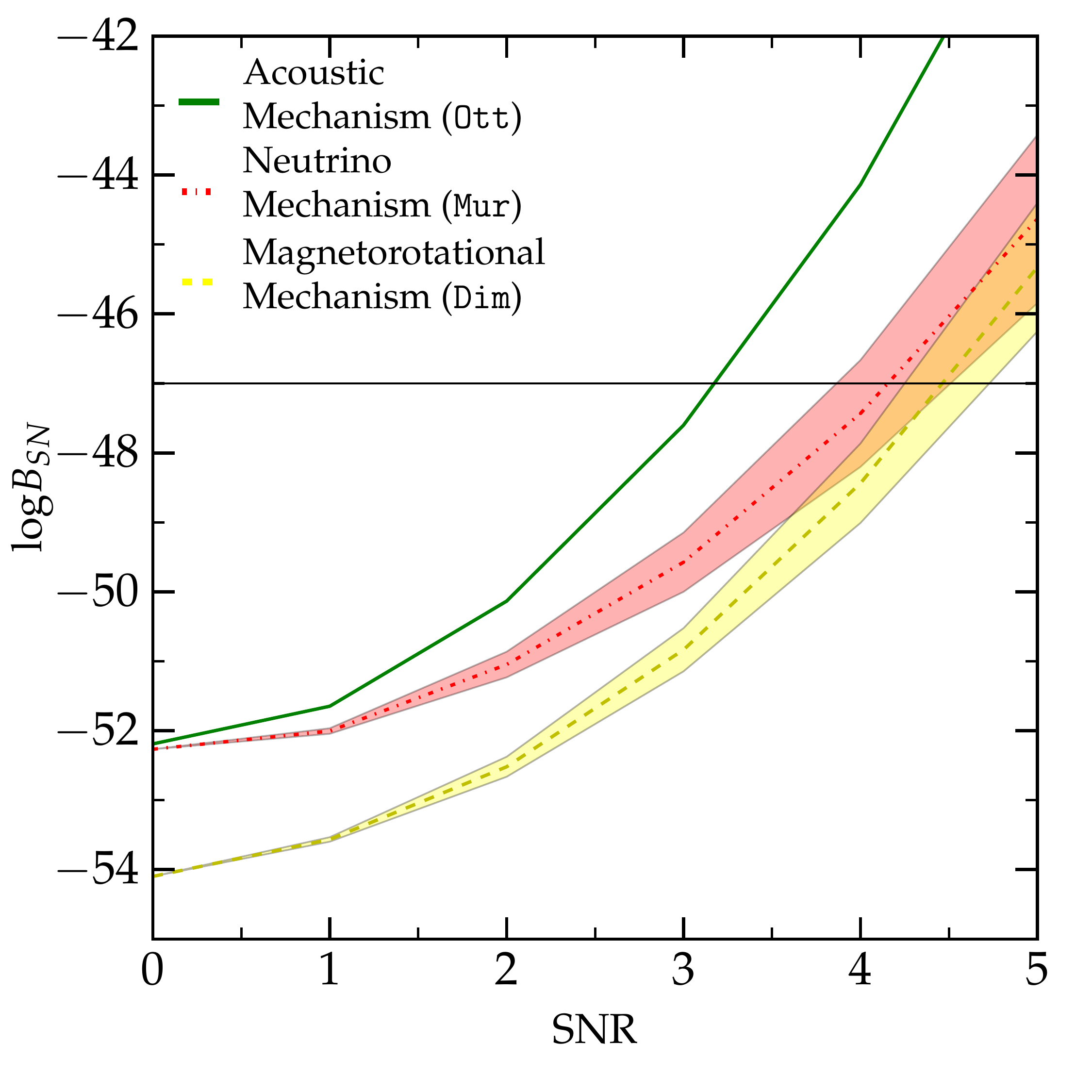}
\caption{Mean $\log{B_{SN}}$ as a function of signal-to-noise ratio
  (SNR; Eq.~\ref{eq:SNR}) for 5 representative waveforms from the
  \cat{Mur}, \cat{Ott} and \cat{Dim} catalogs using 7 Principal
  Components (PC).  The shaded areas represent the standard error in
  the mean value of $\log{B_{SN}}$ for each waveform catalog computed
  as $\sigma = \pm N^{-1} (\Sigma_i (\bar{x} - x_i)^2)^{1/2}$, where
  $\bar{x}$ is the mean and $x_i$ are the individual SNRs and $N$ is
  the number of waveforms. Values of $\log{B_{SN}}$ below $-47$ in the
      7-PC ase and below $-21$ in the 3-PC case indicate that the
      algorithm considers it more likely that there is no signal
      detectable in the noise.  Table~\ref{tab:SNRs} summarizes
      numerical results for the minimum SNR for which $\log{B_{SN}} \ge
      -47$.}
\label{fig:SNRs7PCs}
\end{figure}

\subsection{Signal vs.\ Noise}
\label{sec:svn}

\begin{table}[bt]
\begin{center}
\caption{Minimum SNR at which the injected waveform will give a $\log{B_{SN}}$
    above the threshold defined in \S\ref{sec:purenoise}. This is shown
  for 5 representative waveforms from the \cat{Dim}, \cat{Mur}, and
  \cat{Ott} catalogs, which we take to be representative of the
  magnetorotational, neutrino, and acoustic mechanisms,
  respectively. We provide results from SMEE runs with 3 and 7 PCs. }
\begin{tabular}{l|rr}
\multicolumn{1}{c|}{Waveform}
&\multicolumn{2}{c}{Minimum SNR}\\
\multicolumn{1}{c|}{Name}& 3 PCs & 7 PCs \\
\hline
\cat{Dim} \cite{dimmelmeier:08}&    &     \\
\phantom{0}s11a1o01\_Shen    & 13& 13      \\
\phantom{0}s11a3o09\_Shen    & 4 & 4    \\
\phantom{0}s20a3o05\_LS     & 5 & 5     \\
\phantom{0}s40a3o07\_LS      & 4 & 4   \\
\phantom{0}s40a3o13\_Shen   & 4 & 4   \\
\hline \hline
\cat{Mur} \cite{murphy:09}  &    &      \\
\phantom{0}20\_3.8     & 9 & 4  \\
\phantom{0}40\_10.0   & 4 & 4   \\
\phantom{0}40\_13.0    & 21 & 15 \\
\phantom{0}15\_3.2     & 10 & 9  \\
\phantom{0}15\_4.0    & 4  &  4 \\
\hline \hline
\cat{Ott} \cite{ott:09}  &  &     \\
\phantom{0}nomoto13     & 143 & 4  \\
\phantom{0}nomoto15     & 45 & 4  \\
\phantom{0}s15.0WHW02  & 4 &   4 \\
\phantom{0}s20.0WHW02  & 4 & 4 \\
\phantom{0}s25.0WHW02  & 4 &  4 \\
\hline \hline 
\end{tabular}
\label{tab:SNRs}
\end{center}
\end{table}

\begin{table*}[t!]
\caption{$\log B_{SN}$ with respect to the \cat{Dim}, \cat{Mur}, and
  \cat{Ott} PCs computed for representative injected waveforms from
  the \cat{Dim}, \cat{Mur}, and \cat{Ott} catalogs, which we take to
  be representative of the magnetorotational, neutrino, and acoustic
  explosion mechanism, respectively. Results for source distances of
  $0.2$, $2$, and $10\,\mathrm{kpc}$ are given. Each table entry shows
  the results for $3$ PCs (to the left of the vertical divider
  $|$) and the 7-PC result (to the right of the divider).
  $\log B_{SN} < -47 $ when 7 PCs are used and $\log
    B_{SN} < -21 $ when 3 PCs are used indicates that the injected
    signal is likely consistent with noise while larger values
    suggests that the signal belongs to the signal model whose PCs
    were used in the analysis.}
\begin{center}
\begin{tabular}{l|ccc|cc@{\hspace{-0.2em}}c|c@{\hspace{-0.2em}}c@{\hspace{-0.1em}}c}
\multicolumn{1}{c|}{Waveform}
&\multicolumn{3}{c|}{$\log{B_{SN}}$}
&\multicolumn{3}{c|}{$\log{B_{SN}}$}
&\multicolumn{3}{c}{$\log{B_{SN}}$}\\
\multicolumn{1}{c|}{Name}
&\multicolumn{3}{c|}{\cat{Dim} PCs}
&\multicolumn{3}{c|}{\cat{Mur} PCs}
&\multicolumn{3}{c}{\cat{Ott} PCs}\\
&\multicolumn{1}{c}{0.2 kpc}
&\multicolumn{1}{c}{2 kpc} 
&\multicolumn{1}{c|}{10 kpc}
&\multicolumn{1}{c}{0.2 kpc}
&\multicolumn{1}{c}{2 kpc} 
&\multicolumn{1}{c|}{10 kpc}
&\multicolumn{1}{c}{0.2 kpc}
&\multicolumn{1}{c}{2 kpc}
&\multicolumn{1}{c}{10 kpc}\\
\hline
\cat{Dim} \cite{dimmelmeier:08} & & & & & &\\ 
\phantom{0}s11a1o01\_Shen & 38982$|$48686 &  364$|$430 &  -10$|$-34  & \phantom{0}9178$|$25647 &   \phantom{0}65$|$203 &  -22$|$-42  &  \phantom{0}-25$|$222\phantom{-} &  -27$|$-51 &  -26$|$-52 \\ 
\phantom{0}s11a3o09\_Shen &  $5\!\!\times\!\!10^6$$|$$5\!\!\times\!\!10^6$ & 52403$|$54505 & 2071$|$2129  & 36170$|$58531 &  335$|$532 &  -11$|$-29  &   \phantom{00}23$|$9191 &  -26$|$39\phantom{-} &  -26$|$-49 \\ 
\phantom{0}s20a3o05\_LS   &  $3\!\!\times\!\!10^6$$|$$3\!\!\times\!\!10^6$ & 25738$|$26859 & 1005$|$1023  & \phantom{0}4139$|$29285 &   \phantom{0}15$|$240 &  -24$|$-40  &   \phantom{000}1$|$2242 &  -26$|$-31 &  -26$|$-51 \\ 
\phantom{0}s40a3o07\_LS   &  $1\!\!\times\!\!10^7$$|$$1\!\!\times\!\!10^7$ &  $1\!\!\times\!\!10^5$$|$$1\!\!\times\!\!10^5$ & 4737$|$5411  & 47029$|$$1\!\!\times\!\!10^5$ &  444$|$951 &   \phantom{0}-7$|$-12  &  \phantom{00}519$|$56978 &  \phantom{0}-21$|$517\phantom{-} &  -26$|$-29 \\ 
\phantom{0}s40a3o13\_Shen &  $2\!\!\times\!\!10^7$$|$$2\!\!\times\!\!10^8$ &$2\!\!\times\!\!10^5$$|$$2\!\!\times\!\!10^5$ & 8174$|$8302  & $2\!\!\times\!\!10^5$$|$$4\!\!\times\!\!10^5$ & 1830$|$3968 &   \phantom{0}48$|$109  &  \phantom{00}160$|$19781 &  \phantom{0}-25$|$145\phantom{-} &  -26$|$-44 \\  \hline \hline
\cat{Mur} \cite{murphy:09} &&&&&&\\
\phantom{0}20\_3.8  & \phantom{0}4205$|$10631 &   16$|$50 &  -24$|$-50  & 59264$|$$4\!\!\times\!\!10^5$ &  \phantom{0}566$|$3775 &   \phantom{00}-2$|$101\phantom{-}  &  -22$|$-5\phantom{0} &  -27$|$-53 &  -26$|$-52 \\ 
\phantom{0}40\_10.0 &  \phantom{0}280$|$2362 &  -23$|$-33 &  -26$|$-53  &$4\!\!\times\!\!10^5$$|$$5\!\!\times\!\!10^5$ & 4198$|$4981 &  143$|$149  &  -26$|$-29 &  -27$|$-54 &  -26$|$-52 \\ 
\phantom{0}40\_13.0 &  365$|$819 &  -23$|$-48 &  -26$|$-54  & 3874$|$7591 &   12$|$23 &  -24$|$-49  &  -27$|$-42 &  -27$|$-54 &  -26$|$-52 \\ 
\phantom{0}15\_3.2  & \phantom{0}7450$|$10699 &   48$|$51 &  -23$|$-50  & 23744$|$29439 &  211$|$241 &  -16$|$-41  &  -25$|$-12 &  -27$|$-53 &  -26$|$-52 \\ 
\phantom{0}15\_4.0  & 2030$|$7658 &   \phantom{0}-6$|$20\phantom{-} &  -25$|$-51  & $8\!\!\times\!\!10^5$$|$$8\!\!\times\!\!10^5$ & 7672$|$7812 &  282$|$262  &  -12$|$-22 &  -26$|$-53 &  -26$|$-52 \\  \hline \hline
\cat{Ott} \cite{ott:09} &&&&&&\\
\phantom{0}nomoto13 & 3731$|$6155 &   11$|$4\phantom{0} &  -24$|$-51 &  132$|$313 &  -25$|$-50 &  -26$|$-52  &  \phantom{00}544$|$$1\!\!\times\!\!10^6$&  \phantom{000}-21$|$11893\phantom{-} &  \phantom{0}-26$|$426\phantom{-} \\ 
\phantom{0}nomoto15 & 1520$|$2428 &  -11$|$-32 &  -25$|$-53  &   \phantom{0}54$|$146 &  -26$|$-52 &  -26$|$-52  & 32145$|$$7\!\!\times\!\!10^6$&  \phantom{00}295$|$65635 &  \phantom{00}-13$|$2575\phantom{-} \\ 
\phantom{0}s15.0WHW02 &  \phantom{0}578$|$2417 &  -20$|$-33 &  -26$|$-53  & 1931$|$2601 &  \phantom{0}-7$|$-27 &  -25$|$-51  & $6\!\!\times\!\!10^7$$|$$6\!\!\times\!\!10^7$ & $6\!\!\times\!\!10^5$$|$$6\!\!\times\!\!10^5$ & 24653$|$24655 \\ 
\phantom{0}s20.0WHW02 &  \phantom{0}622$|$2895 &  -20$|$-28 &  -26$|$-53  & \phantom{0}-19$|$587\phantom{-} &  -27$|$-47 &  -26$|$-52  & $4\!\!\times\!\!10^7$$|$$4\!\!\times\!\!10^7$ & $4\!\!\times\!\!10^5$$|$$4\!\!\times\!\!10^5$ & 17975$|$17957 \\ 
\phantom{0}s25.0WHW02 & 2343$|$5963 &   -3$|$3\phantom{-} &  -25$|$-51  & 1918$|$2969 &   \phantom{0}-7$|$-24 &  -25$|$-51  & $1\!\!\times\!\!10^8$$|$$1\!\!\times\!\!10^8$ & $1\!\!\times\!\!10^6$$|$$1\!\!\times\!\!10^6$ & 45853$|$45829 \\    \hline \hline
\end{tabular}
\end{center}
\label{tab:mech1}
\end{table*}

The minimal GW signal strength required for SMEE to be able to select
the core-collapse supernova mechanism is an important question.  The
primary prerequisite for an incident GW signal to be useful for model
selection is that SMEE can distinguish it from detector noise, i.e.,
we must find the minimum signal strength (i.e., SNR) so that
$\log{B_{SN}} > -47$ (when 7 PCs are used; Eq.~\ref{eq:sigvsnoise} and
\S\ref{sec:purenoise}).

In order to determine the range of minimum SNR required across and
within core-collapse supernova GW signal types, we draw 5
representative waveforms from the \cat{Dim}, \cat{Mur}, and \cat{Ott}
catalogs and run them through SMEE at varying SNR, using 7 PCs
generated from the catalog to which each injected waveform
belongs. The result of this exercise is shown in
Fig.~\ref{fig:SNRs7PCs} and summarized in
Tab.~\ref{tab:SNRs}. Generally, an SNR $\gtrsim 4-5$ is required for
SMEE to find $\log{B_{SN}} > -47$ in the idealized setting that
we consider here. In a real fully blind search, unknown arrival times
and non-Gaussianity of real detector noise will generally require an
SNR in excess of $8$ for a detection statement (e.g., \cite{abadie:10pop}).

In Fig.~\ref{fig:SNRs7PCs}, the waveforms associated with the acoustic
mechanism (\cat{Ott} catalog) require the smallest SNR, followed by those of
the neutrino mechanism (\cat{Mur} catalog) and the magnetorotational mechanism (\cat{Dim} catalog). 
This hierarchy in minimum SNR,
however, is not fundamental, but a consequence of the fact that we
have chosen to carry out this test using 7 PCs for each waveform
catalog. Since the \cat{Ott} catalog comprises only 7 waveforms, the set of
7 PCs completely spans it and allows perfect reconstruction, 
maximizing $p(D|M_s)$ (Eq.~\ref{eq:like}). In the case of less than
perfect knowledge of the signal, the minimum SNR will always be
greater. This is why the \cat{Dim} and \cat{Mur} catalogs, which have many more
than 7 waveforms, require larger minimum SNR than the \cat{Ott} waveforms.

The situation is somewhat different, when we re-calculate
$\log{B_{SN}}$ using only 3 PCs. The \cat{Ott} catalog is very small
and rather diverse in the time domain. Its first few PCs do not
efficiently span the catalog and, when only the first 3 PCs are used,
the minimum SNRs for waveforms poorly reconstructed with these PCs
increases dramatically, as shown for the \cat{Ott} waveforms nomoto13
and nomoto15 in Tab.~\ref{tab:SNRs}.  Some of the \cat{Mur} waveforms
also exhibit increased minimum SNRs, indicating that there is
significant time-domain variation that is not captured by the first 3
PCs. The large \cat{Dim} catalog, on the other hand, is very
efficiently spanned already by its first few PCs, as previously
pointed out by \cite{heng:09,roever:09}, and the minimum SNR for
waveforms from this catalog remains practically unchanged when going
from 7 to 3 PCs in our analysis.

\subsection{Determining the Core-Collapse Supernova Explosion Mechanism}
\label{sec:mech}

The basic assumption of this study is that the neutrino,
magnetorotational, and acoustic core-collapse supernova explosion
mechanism have robustly distinct GW signatures. In this section, we
test this assumption by injecting waveforms into simulated noise and
running SMEE on the data using PCs of waveform catalogs representative
of the neutrino, magnetorotational, and acoustic mechanisms. If our
assumption is correct and the GW signatures of these mechanisms are
truly distinct, then SMEE should (\emph{i}) yield the largest value of
$\log{B_{SN}}$ when the set of PCs is used that corresponds to the
mechanism the waveform is representative of, and, (\emph{ii})
$\log{B_{ij}}$ (Eq.~\ref{eq:bayes}) should be positive (and larger
than $\sim$5; see \S\ref{sec:purenoise}) if the injected waveform is
most consistent with mechanism $i$, negative if it is most consistent
with mechanism $j$, and near zero if the result is inconclusive.

We carry out our SMEE calculations for events located at
$0.2\,\mathrm{kpc}$, $2\,\mathrm{kpc}$, and $10\,\mathrm{kpc}$ and
with $3$ and $7$ PCs. Betelgeuse ($\alpha$ Orionis), a red supergiant
star of $15-20\,M_\odot$ \cite{smith:09}, is located $197 \pm
45\,\mathrm{pc}$ from Earth \cite{harper:08} and will eventually
explode as a Type II-P core-collapse supernova. Hence, studying a
potential event at $0.2\,\mathrm{kpc}$ will tell us what we can learn
from the observation of GWs from Betelgeuse's collapse and supernova.
$2\,\mathrm{kpc}$ is still nearby on the galactic
scale, but the Galactic volume out to this radius already contains
hundreds of supergiants, one of which may make the next galactic
supernova \cite{hohle:10}. Finally, $10\,\mathrm{kpc}$ is the fiducial
Galactic distance scale and we consider it to state what could be
inferred throughout the Milky Way. As in Sec.~\ref{sec:svn}, we carry
out SMEE runs with 3 and 7 PCs to study the sensitivity of the results
on the amount of knowledge about the injected waveform we grant 
SMEE.

Table~\ref{tab:mech1} summarizes the results from $\log{B_{SN}}$
calculations for five representative waveforms from the \cat{Mur}
(neutrino mechanism), \cat{Dim} (magnetorotational mechanism), and
\cat{Ott} (acoustic mechanism) catalogs. The larger the value of
$\log{B_{SN}}$, the greater the confidence that the data contain a
signal consistent with the employed set of PCs. 
  $\log{B_{SN}} \le -47$ (for the 7-PC case; $-21$ in the 3-PC case)
indicates that the signal is more consistent with noise.  The results
show that it is indeed possible to clearly associate any injected
waveform with its catalog and, thus, select the explosion mechanism
for a core-collapse supernova out to at least
$2\,\mathrm{kpc}$. However, at the galactic scale
($10\,\mathrm{kpc}$), a significant fraction of waveforms
  representative of the neutrino mechanism, due to their intrinsically
  low GW amplitudes, can be told apart from noise only marginally or
  are even most consistent with noise. Typical magnetorotational
explosions and explosions driven by the acoustic mechanism are still
clearly identifiable at $10\,\mathrm{kpc}$. However, when the reduced
set of 3 PCs is used, the \cat{Ott} catalog, which is not
  well spanned by only 3 PCs, suffers most and two out of the five
  representative \cat{Ott} waveforms are not or only marginally
  identifiable with the 3 first \cat{Ott} PCs at $10\,\mathrm{kpc}$.

Provided that there is confidence that a signal has been detected, we
can compute $\log B_{ij}$ (Eq.~\ref{eq:bayes}) to study if the signal
is more likely to be consistent with mechanism $i$ or mechanism $j$.
Since we know $\log B_{SN}$ for all signal models, we can simply
compute $\log B_{ij} = \log B_{iN} - \log B_{jN}$ from the data for
representative waveforms provided in
Tab.~\ref{tab:mech1}. We use $7$ PCs for these
  calculations.

\begin{figure*}[htb]
\includegraphics[width=0.32\textwidth]{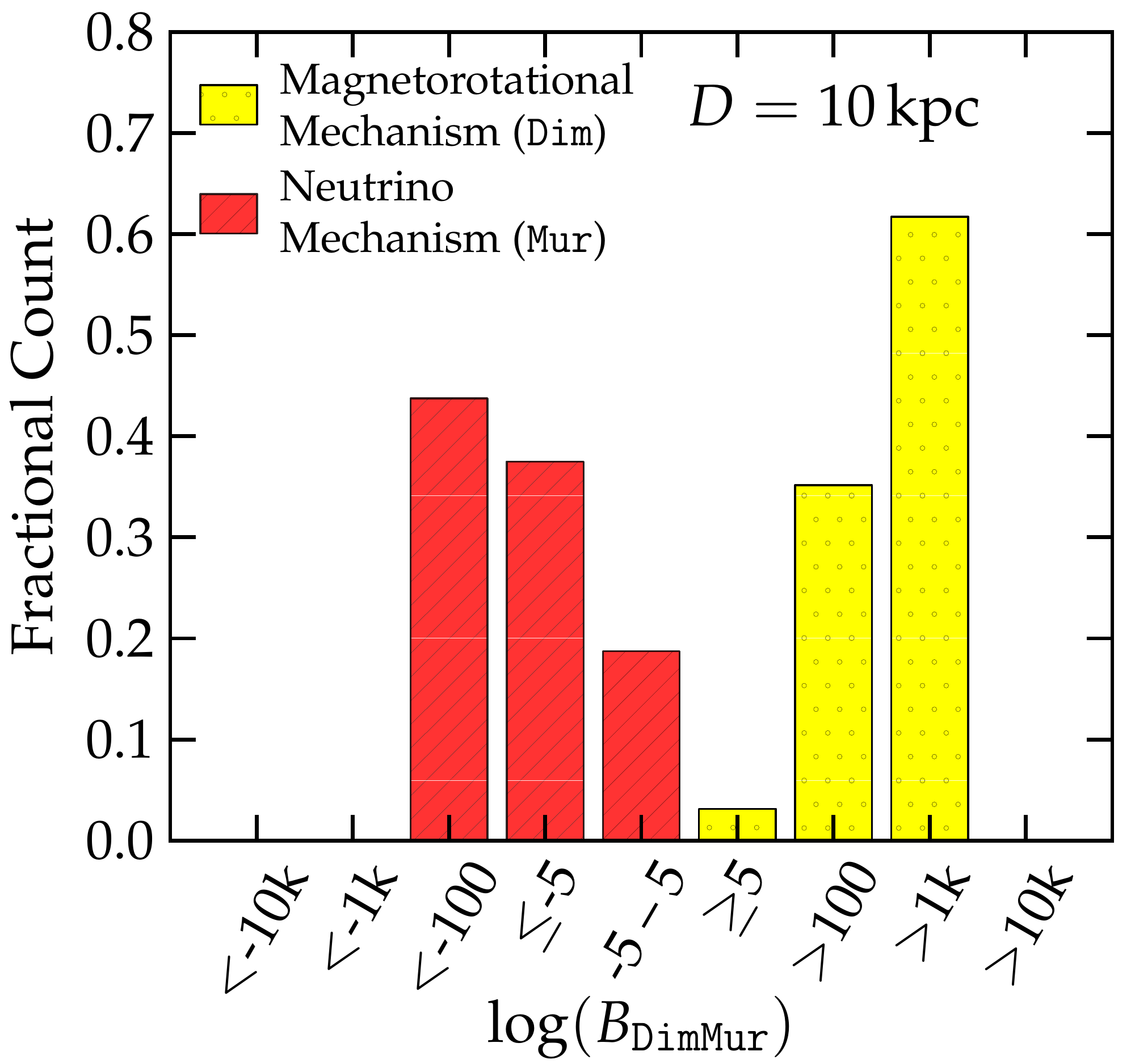}
\includegraphics[width=0.32\textwidth]{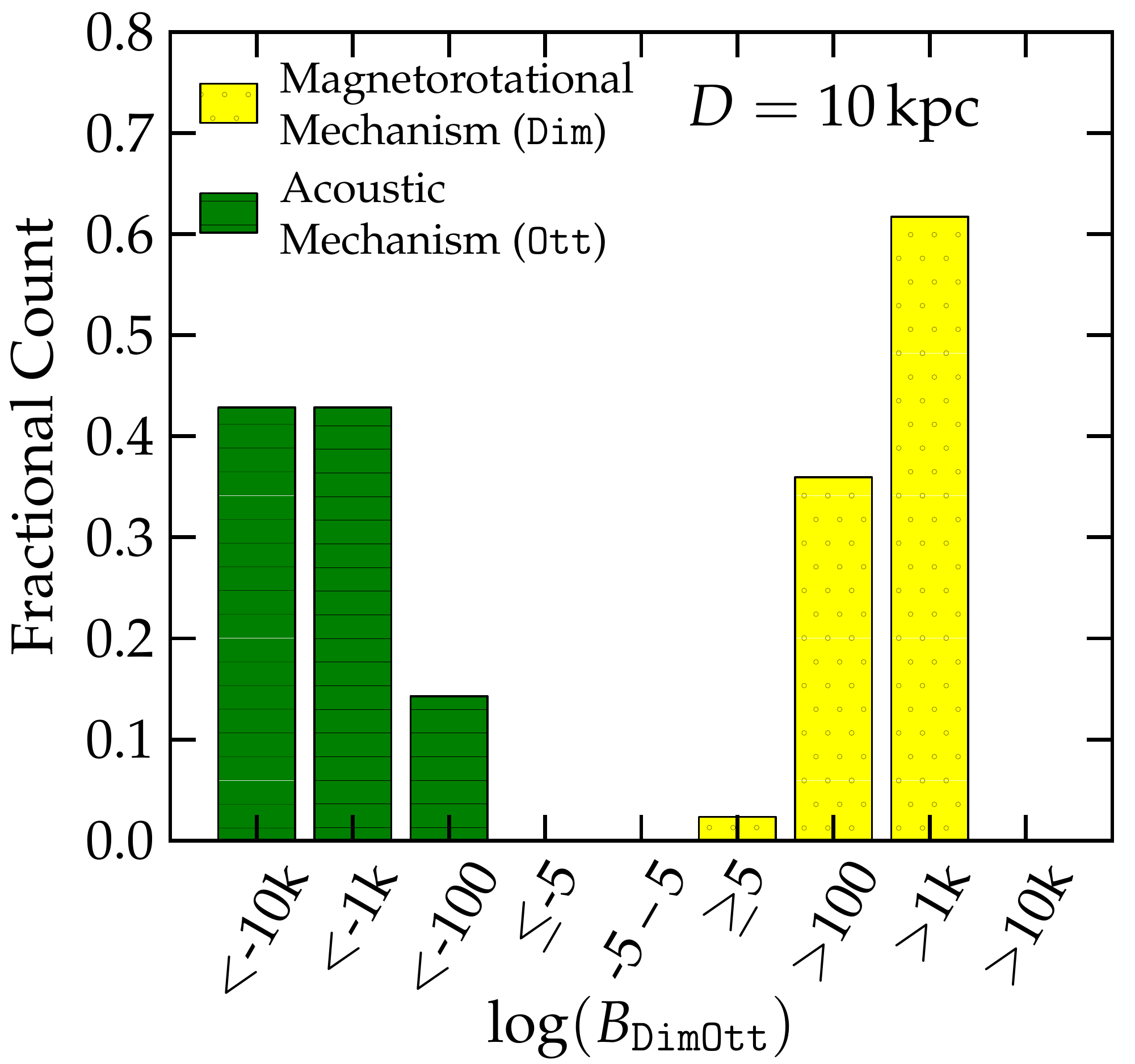}
\includegraphics[width=0.32\textwidth]{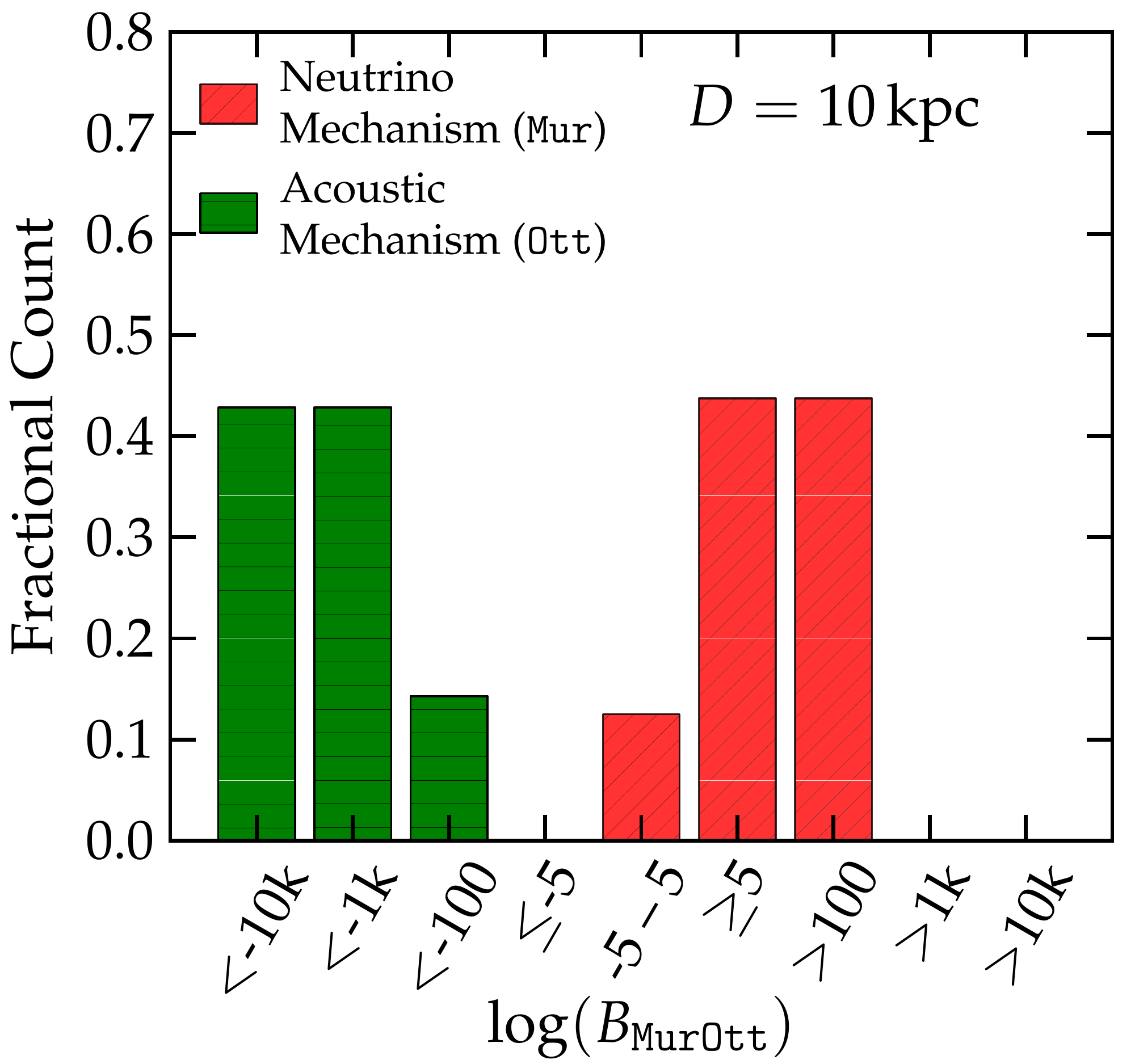}
\caption{Histograms describing the outcome of signal model comparisons
  by means of the Bayes factors $\log B_{ij} = \log p(D|M_i) - \log
  p(D|M_j)$, where $i \ne j$ and $M_i$ and $M_j$ are signal models
  described by the \cat{Dim} (magnetorotational mechanism), \cat{Mur}
  (neutrino mechanism), and \cat{Ott} (acoustic mechanism) waveform
  catalogs.  The Bayes factors are computed with $7$ PCs and for a
  source distance of $10\,\mathrm{kpc}$. A positive value $\log
  B_{ij}$ indicates that the injected waveform most likely belongs to
  model $M_i$, while a negative value suggest that model $M_j$ is the
  more probable explanation. The bars are color-coded according to the
  type of injected waveform. The results are binned into ranges of
  varying size from $< -10000$ to $> 10000$ and the height of the bars
  indicates what fraction of the waveforms of a given catalog falls
  into a given bin of $\log B_{ij}$. We consider the range of $(-5,5)$
  of $\log B_{ij}$ as inconclusive evidence (see \S\ref{sec:purenoise}).}
\label{fig:histall}
\end{figure*}

In Fig.~\ref{fig:histall}, we show results of injection studies of
\emph{all} waveforms from the \cat{Dim}, \cat{Mur}, and \cat{Ott}
catalogs run through SMEE and analyzed with the \cat{Dim}, \cat{Mur},
and \cat{Ott} PCs at a source distance of $10\,\mathrm{kpc}$. The full
numerical results on whose basis Fig.~\ref{fig:histall} was generated
are available on-line~\cite{smee1web}.  The left panel depicts the
$\log B_{\tt DimMur}$ result for injected waveforms from the \cat{Dim}
and \cat{Mur} catalogs, that we take to be representative of the
magnetorotational and neutrino mechanism, respectively. Even at
$10\,\mathrm{kpc}$ the vast majority of waveforms characteristic for
magnetorotational explosions are clearly identified as belonging to
this mechanism. For the neutrino mechanism, the evidence is generally
significantly weaker and only $\sim$$44$\% of the \cat{Mur} waveforms
are identified with $\log B_{\tt DimMur} < -100$ and none have $\log
B_{\tt DimMur} < -1000$, while $\sim$19\% are in the inconclusive
regime of $-5 < \log B_{\tt DimMur} < 5$.

In the center panel of Fig.~\ref{fig:histall}, we show results for
$\log B_{\tt DimOtt}$ for injected waveforms corresponding to the
magnetorotational (\cat{Dim}) and the acoustic (\cat{Ott}) mechanism.
The case is clear cut and most waveforms are correctly identified as
most likely belonging to their respective catalog/mechanism.  Finally,
the right panel of Fig.~\ref{fig:histall} presents $\log B_{\tt
  MurOtt}$ for waveforms representative of the neutrino (\cat{Mur})
and acoustic (\cat{Ott}) mechanism. As in the previous panel, SMEE
associates the waveforms corresponding to the acoustic mechanism with
high confidence to the \cat{Ott} catalog.  The evidence suggesting
correct association of the neutrino mechanism waveforms is
considerably less strong, but $\log B_{\tt MurOtt}$ is still 
conclusive for $\sim$88\% of the \cat{Mur} waveforms.

\begin{figure*}[htb]
\includegraphics[width=0.32\textwidth]{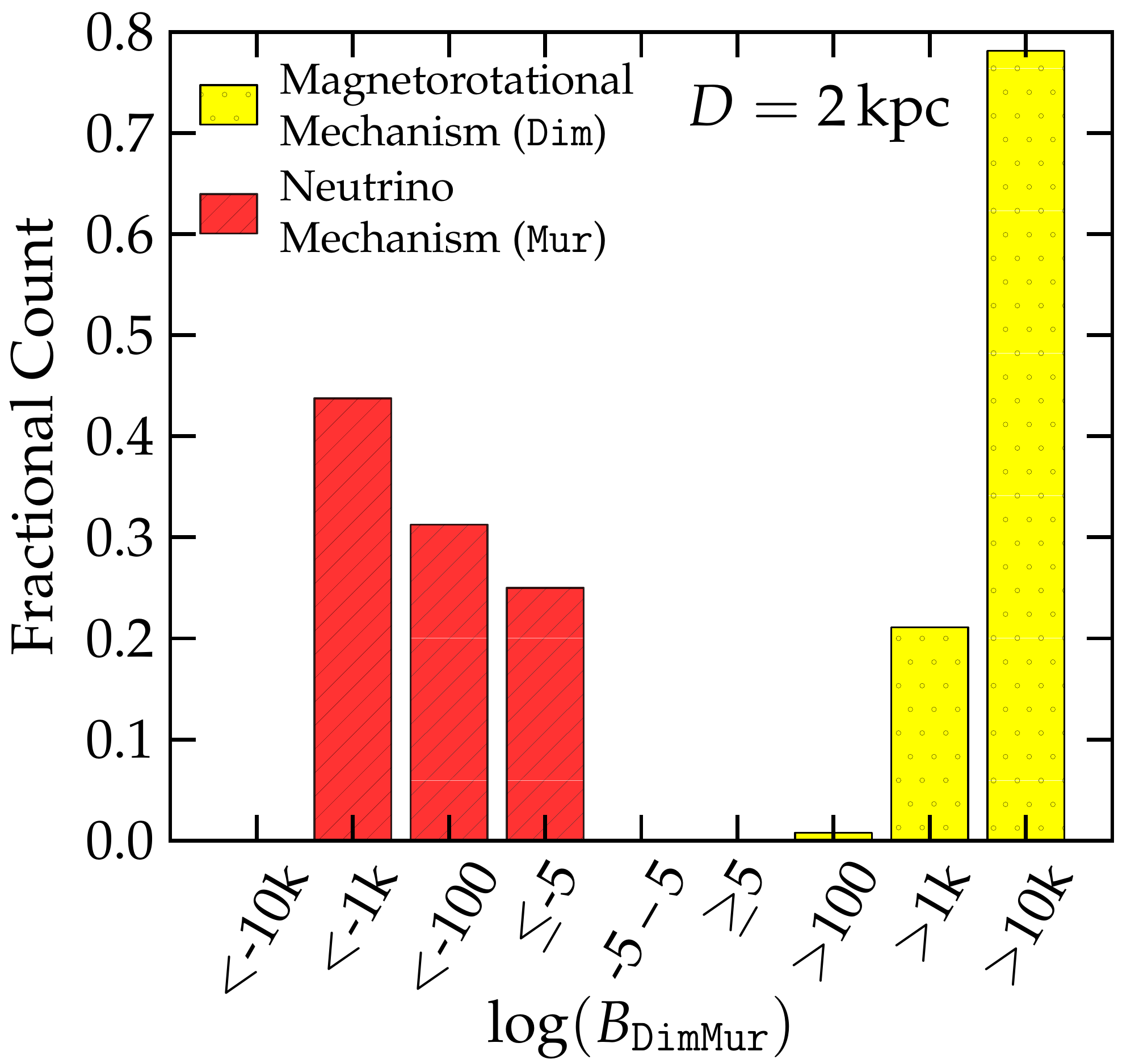}
\includegraphics[width=0.32\textwidth]{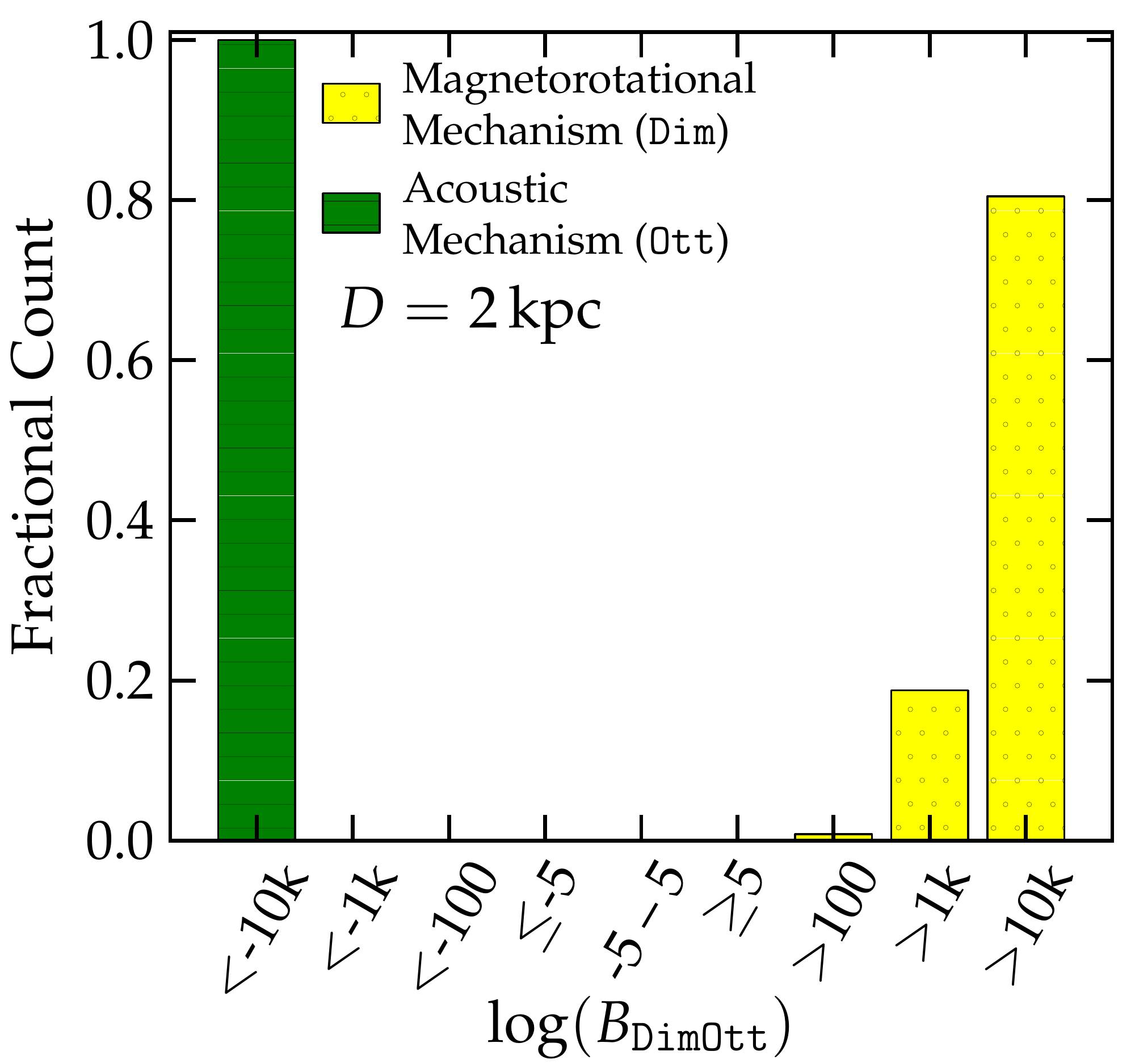}
\includegraphics[width=0.32\textwidth]{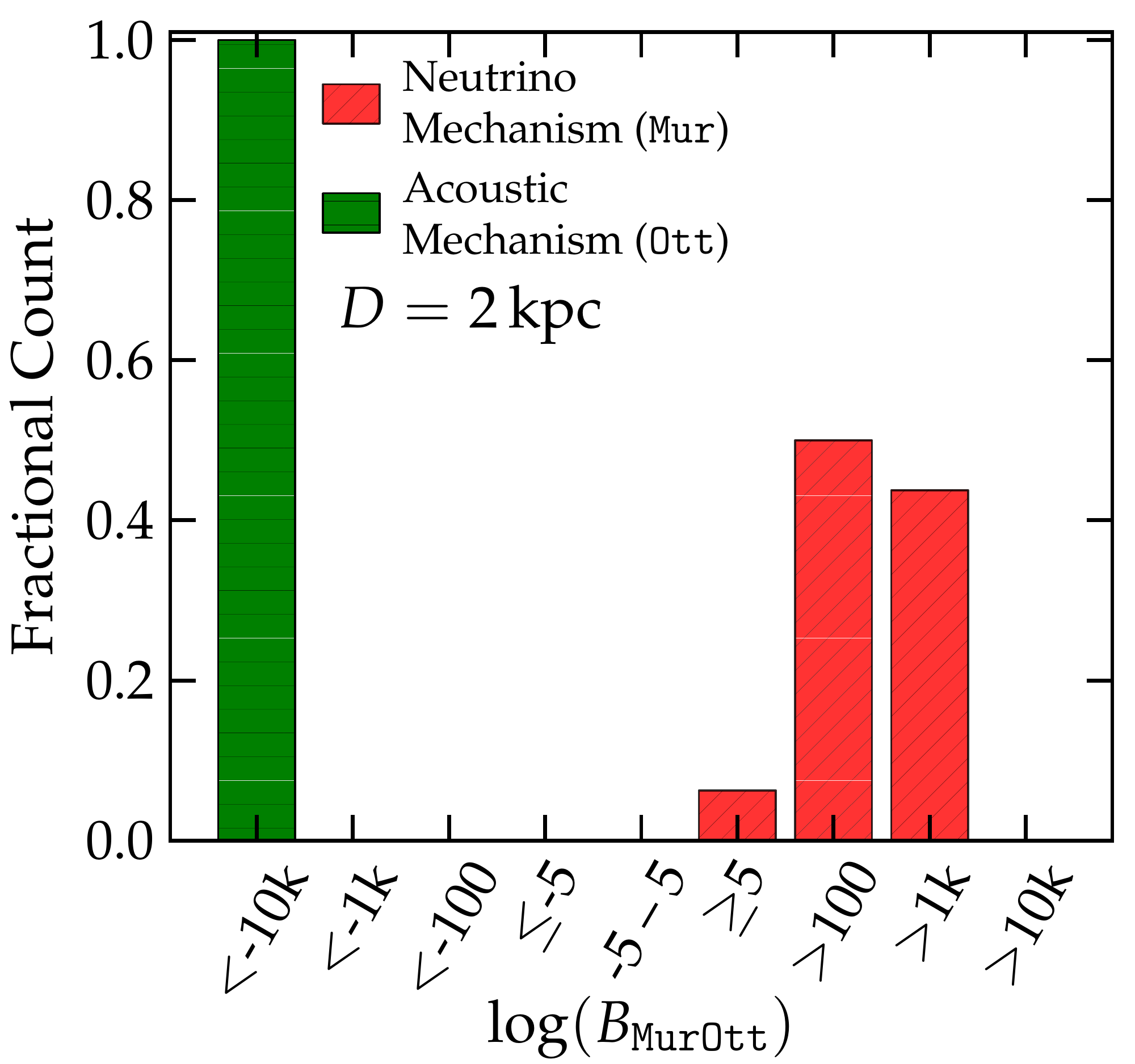}
\caption{Same as Fig.~\ref{fig:histall}, but computed for a source distance
of $2\,\mathrm{kpc}$.}
\label{fig:histall_2kpc}
\end{figure*}

Figure~\ref{fig:histall_2kpc} shows the results for $\log B_{\tt
  DimMur}$, $\log B_{\tt DimOtt}$, and $\log B_{\tt MurOtt}$ obtained
by SMEE with $7$ PCs at a source distance of $2\,\mathrm{kpc}$. Here,
all acoustic mechanism waveforms (\cat{Ott} catalog), all magnetorotational
mechanism waveforms (\cat{Dim} call), and all neutrino mechanism waveforms
(\cat{Mur} catalog) are correctly identified as belonging to their
respective catalog and explosion mechanism.

\begin{figure*}[tb]
\centering
\includegraphics[width=0.33\textwidth]{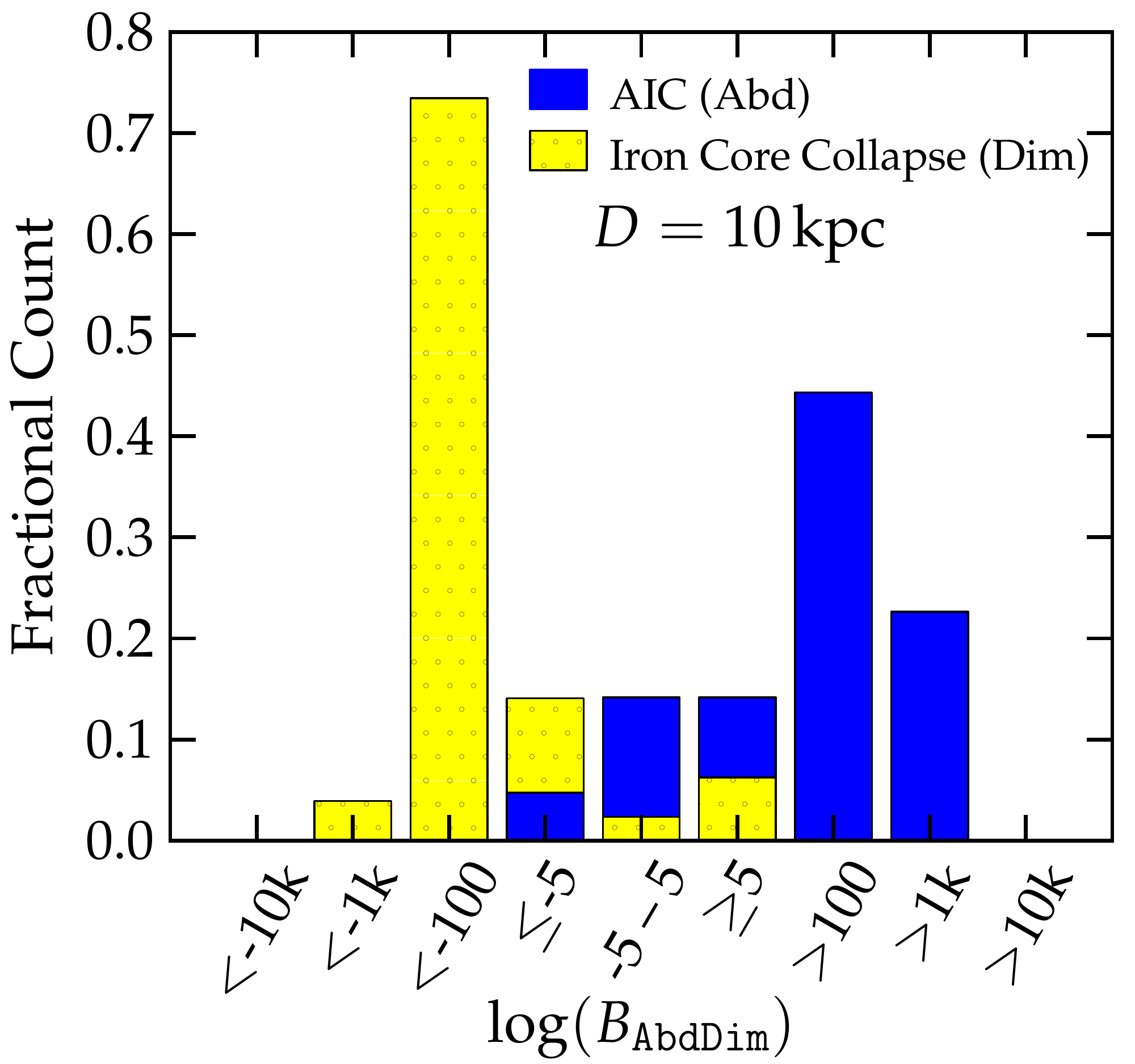}
\hspace*{0.35cm}
\includegraphics[width=0.33\textwidth]{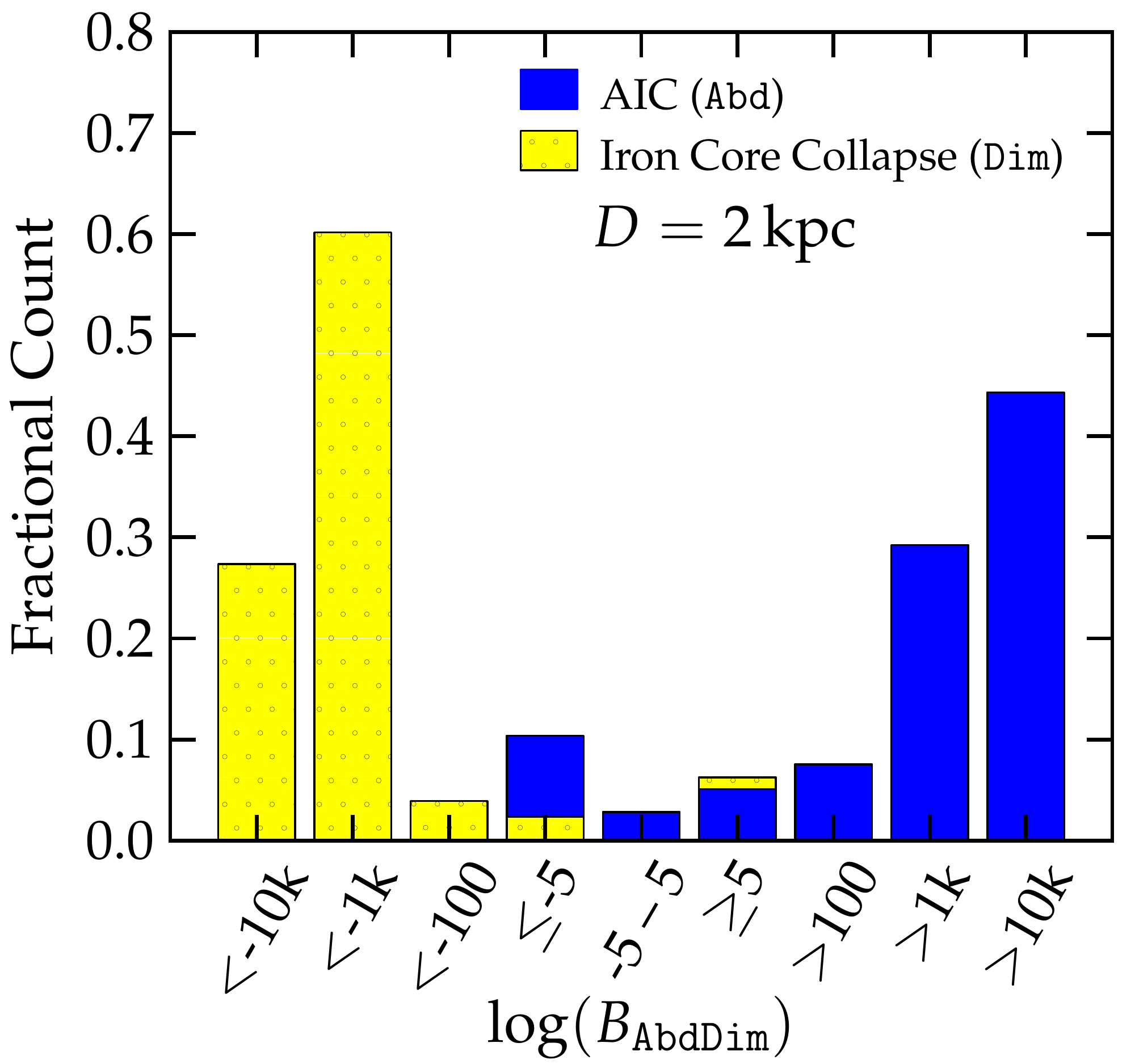}
\caption{Outcome of the SMEE analysis of injected rotating iron core
  collapse (\cat{Dim} catalog) and rotating accretion-induced
  collapse (AIC, \cat{Abd} catalog) waveforms. The left panel
  shows results for a source distance of $10\,\mathrm{kpc}$ and the
  right panel depicts the results for a distance of $2\,\mathrm{kpc}$.
  The Bayes factors $\log B_{\tt AbdDim}$ are computed on the basis
  of 7 PCs from the \cat{Abd} and \cat{Dim} catalog.  A positive value of $\log
  B_{\tt AbdDim}$ indicates that an injected waveform is most
  likely associated with rotating AIC and a negative value suggests it
  to be more consistent with rotating iron core collapse. The results
  are binned into ranges of varying size from $< -10000$ to $> 10000$
  and the height of the color-coded bars indicates what fraction of
  the waveforms of a given catalog falls into a given bin of $\log
  B_{\tt AbdDim}$.  We consider the range of $(-5,5)$
  of $\log B_{ij}$ as inconclusive evidence (see \S\ref{sec:purenoise}).}
\label{fig:AD}
\end{figure*}

\begin{table*}[t!]
\begin{center}
  \caption{$\log B_{SN}$ for gravitational waveforms that were not
    included in the catalogs used for PC computation. The
    \cat{DimExtra}, \cat{Sch}, \cat{OttExtra}, and \cat{Yak} waveforms
    are discussed in \S\ref{sec:snmech}. The \cat{MurRem} waveforms are three
    randomly selected waveforms from the \cat{Mur} catalog that were removed
    before re-computation of the \cat{Mur} PCs.
    Results are shown for source distances of $0.2\,\mathrm{kpc}$,
    $2\,\mathrm{kpc}$, and $10\,\mathrm{kpc}$ and for evaluations
    using 3 PCs (to the left of the vertical divider $|$) and 7 PCs (to 
    the right of the divider).
    Larger values indicate stronger evidence that the
    waveform is matched to the model catalog from which the PCs were
    constructed. $\log B_{SN} < -47 $ when 7 PCs are used and $\log
    B_{SN} < -21 $ when 3 PCs are used indicates that the injected
    signal is likely consistent with noise while larger values
    suggests that the signal belongs to the signal model whose PCs
    were used in the analysis.}
\begin{tabular}{l|ccc|cc@{\hspace{-0.2em}}c|c@{\hspace{-0.2em}}c@{\hspace{-0.3em}}c}
\multicolumn{1}{c|}{Waveform}
&\multicolumn{3}{c|}{$\log{B_{SN}}$}
&\multicolumn{3}{c|}{$\log{B_{SN}}$}
&\multicolumn{3}{c}{$\log{B_{SN}}$}\\
\multicolumn{1}{c|}{Name}
&\multicolumn{3}{c|}{\cat{Dim} PCs}
&\multicolumn{3}{c|}{\cat{Mur} PCs}
&\multicolumn{3}{c}{\cat{Ott} PCs}\\
&\multicolumn{1}{c}{0.2 kpc}
&\multicolumn{1}{c}{2 kpc} 
&\multicolumn{1}{c|}{10 kpc}
&\multicolumn{1}{c}{0.2 kpc}
&\multicolumn{1}{c}{2 kpc} 
&\multicolumn{1}{c|}{10 kpc}
&\multicolumn{1}{c}{0.2 kpc}
&\multicolumn{1}{c}{2 kpc}
&\multicolumn{1}{c}{10 kpc}\\
 \hline
\cat{DimExtra} \cite{roever:09} & & & & & &\\
\phantom{0}s20a1o05\_shen  & $3\!\!\times\!\!10^6$$|$$3\!\!\times\!\!10^6$ & 28269$|$31142 & 1106$|$1194  & $1\!\!\times\!\!10^5$$|$$2\!\!\times\!\!10^5$ & 1429$|$2020 &   32$|$31  &  \phantom{0}352$|$6515 &  -23$|$12\phantom{0} &  -26$|$-49 \\ 
\phantom{0}s15a1o03\_LS    & $8\!\!\times\!\!10^6$$|$$9\!\!\times\!\!10^6$ & 89606$|$95070 & 3560$|$3751& $2\!\!\times\!\!10^5$$|$$3\!\!\times\!\!10^5$ & 1966$|$2593 &   54$|$54  &\phantom{00}384$|$40334 &  -22$|$350 &  -26$|$-36 \\ 
\phantom{0}s40a1o10\_LS    & $2\!\!\times\!\!10^7$$|$$2\!\!\times\!\!10^7$ & $2\!\!\times\!\!10^5$$|$$2\!\!\times\!\!10^5$ & 6832$|$7564  & $3\!\!\times\!\!10^5$$|$$1\!\!\times\!\!10^6$ &\phantom{0}2624$|$14076 &   \phantom{0}80$|$513  &  \phantom{00}142$|$13089 &  -25$|$78\phantom{-} &  -26$|$-47 \\  \hline 
\cat{Sch} \cite{scheidegger:10b} & & & & & & & &\\
\phantom{0}R1E1CA     & 20861$|$32817 &  182$|$271 &  -17$|$-41  &   \phantom{0}68$|$482 &  -26$|$-48 &  -26$|$-52  &  -27$|$68\phantom{-} &  -27$|$-52 &  -26$|$-52 \\ 
\phantom{0}R1E1CA\_L  & 11631$|$13694 & 90$|$80 & -21$|$-48 &  -25$|$-48 &  -27$|$-54 &  -26$|$-52  &  -27$|$19\phantom{-} &  -27$|$-53 &  -26$|$-52 \\ 
\phantom{0}R1E1DB     & 19285$|$25307 &  167$|$196 &  -18$|$-44  &   \phantom{0}63$|$438 &  -26$|$-49 &  -26$|$-52  &  -26$|$89\phantom{-} &  -27$|$-52 &  -26$|$-52 \\ 
\phantom{0}R1E3CA     & 27629$|$46426 & 250$|$408 & -15$|$-35 &   \phantom{-}1$|$-6 &  -26$|$-53 &  -26$|$-52  &  \phantom{1}-23$|$160\phantom{-} &  -27$|$-52 &  -26$|$-52 \\ 
\phantom{0}R1STCA     & \phantom{0}8304$|$10006 &   57$|$44 &  -23$|$-50  &  114$|$127 &  -25$|$-52 &  -26$|$-52  &  -27$|$-33 &  -27$|$-54 &  -26$|$-52 \\ 
\phantom{0}R2E1AC     & $4\!\!\times\!\!10^5$$|$$4\!\!\times\!\!10^5$ & 3923$|$4273 & 132$|$120 & 1133$|$1602 &  -15$|$-37 &  -25$|$-52  &  \phantom{1}-19$|$991\phantom{-} &  -26$|$-43 &  -26$|$-52 \\ 
\phantom{0}R2E3AC      & $4\!\!\times\!\!10^5$$|$$4\!\!\times\!\!10^5$ & 3566$|$3902 &  118$|$104  & 1151$|$1567 &  -15$|$-38 &  -25$|$-52  &  \phantom{1}-18$|$578\phantom{-} &  -26$|$-47 &  -26$|$-52 \\ 
\phantom{0}R2STAC      & $8\!\!\times\!\!10^5$$|$$8\!\!\times\!\!10^5$ & 7504$|$7785 & 275$|$260 &  \phantom{0}405$|$1301 &  -22$|$-40 &  -26$|$-52  &  \phantom{14}-25$|$1463\phantom{-} &  -27$|$-39 &  -26$|$-52 \\ 
\phantom{0}R3E1AC      & $3\!\!\times\!\!10^6$$|$$4\!\!\times\!\!10^6$ & 28195$|$37420 & 1103$|$1445  & \phantom{0}7418$|$13743 &   48$|$84 &  -23$|$-47  &  \phantom{00}126$|$10015 &  -25$|$47\phantom{-} &  -26$|$-48 \\ 
\phantom{0}R3E1AC\_L   & $2\!\!\times\!\!10^6$$|$$3\!\!\times\!\!10^6$ & 19227$|$25641 & 744$|$974 & 10361$|$11138 &   77$|$58 &  -22$|$-48  &  \phantom{0}103$|$5089 &  -25$|$-2\phantom{0} &  -26$|$-50 \\ 
\phantom{0}R3E1CA      & $2\!\!\times\!\!10^6$$|$$3\!\!\times\!\!10^6$ & 20652$|$30427 &  \phantom{0}801$|$1165  & \phantom{0}7590$|$13598 &   50$|$83 &  -23$|$-47  &  \phantom{0}183$|$7814 &  -24$|$25\phantom{-} &  -26$|$-49 \\ 
\phantom{0}R3E1DB      & $2\!\!\times\!\!10^6$$|$$3\!\!\times\!\!10^6$ & 20722$|$30537 & \phantom{0}804$|$1170 & 10102$|$18050 &   \phantom{0}75$|$127 &  -22$|$-45  &  \phantom{0}176$|$7438 &  -25$|$21\phantom{-} &  -26$|$-49 \\ 
\phantom{0}R3E2AC      & $2\!\!\times\!\!10^6$$|$$3\!\!\times\!\!10^6$& 24203$|$27271 & \phantom{0}943$|$1039  & \phantom{0}4575$|$12516 &   19$|$72 &  -24$|$-47  &   \phantom{00}32$|$7135 &  -26$|$18\phantom{-} &  -26$|$-49 \\ 
\phantom{0}R3E3AC      & $3\!\!\times\!\!10^6$$|$$4\!\!\times\!\!10^6$ & 33975$|$39403 & 1334$|$1524 & 5493$|$8915 &   29$|$36 &  -24$|$-49  &  \phantom{00}107$|$13629 &  -25$|$83\phantom{-} &  -26$|$-47 \\ 
\phantom{0}R3STAC      & $5\!\!\times\!\!10^6$$|$$5\!\!\times\!\!10^6$ & 47277$|$50486 & 1866$|$1968  & 13231$|$17583 &  106$|$122 &  -21$|$-45  &   \phantom{000}70$|$10361 &  -26$|$50\phantom{-} &  -26$|$-48 \\ 
\phantom{0}R4E1AC      & $9\!\!\times\!\!10^6$$|$$1\!\!\times\!\!10^7$ & 87917$|$$1\!\!\times\!\!10^5$& 3492$|$4121 & 15672$|$26725 &  130$|$214 &  -20$|$-42  &  \phantom{00}584$|$39541 &  \phantom{0}-20$|$342\phantom{-} &  -26$|$-36 \\ 
\phantom{0}R4E1CF      & $5\!\!\times\!\!10^7$$|$$6\!\!\times\!\!10^7$ & $5\!\!\times\!\!10^5$$|$$6\!\!\times\!\!10^5$ & 21092$|$22361  & $8\!\!\times\!\!10^5$$|$$3\!\!\times\!\!10^6$ & \phantom{0}7543$|$31753 &  \phantom{0}277$|$1220  & \phantom{00}655$|$$1\!\!\times\!\!10^5$ &  \phantom{01}-20$|$1434\phantom{-} &  -26$|$7\phantom{-0} \\ 
\phantom{0}R4E1EC      & $8\!\!\times\!\!10^6$$|$$8\!\!\times\!\!10^7$ & 75553$|$82557 & 2997$|$3251 & 15653$|$31961 &  130$|$266 &  -20$|$-39  &  \phantom{00}315$|$34696 &  \phantom{0}-23$|$294\phantom{-} &  -26$|$-38 \\ 
\phantom{0}R4E1FC      & $4\!\!\times\!\!10^7$$|$$4\!\!\times\!\!10^7$ & $4\!\!\times\!\!10^4$$|$$4\!\!\times\!\!10^5$ & 16017$|$17730  &$3\!\!\times\!\!10^5$$|$$5\!\!\times\!\!10^5$ & 3140$|$5586 &  101$|$173  &  \phantom{00}643$|$$1\!\!\times\!\!10^5$ &  \phantom{00}-20$|$1253\phantom{-} &  -26$|$0\phantom{0-} \\ 
\phantom{0}R4E1FC\_L   & $8\!\!\times\!\!10^6$$|$$1\!\!\times\!\!10^7$ & 83536$|$97750 & 3317$|$3859 & 16402$|$21202 &  138$|$159 &  -19$|$-44  &  \phantom{00}301$|$33112 &  \phantom{1}-23$|$278\phantom{-} &  -26$|$-39 \\ 
\phantom{0}R4STAC      & $9\!\!\times\!\!10^6$$|$$1\!\!\times\!\!10^7$& 94188$|$$1\!\!\times\!\!10^5$ & 3743$|$5146  & 45510$|$63122 &  429$|$578 &   \phantom{0}-8$|$-27  & \phantom{0}1310$|$37004 &  \phantom{1}-13$|$317\phantom{-} &  -25$|$-37 \\ 
\phantom{0}R5E1AC      & $7\!\!\times\!\!10^6$$|$$8\!\!\times\!\!10^6$ & 70290$|$78739 & 2787$|$3098 & 36378$|$45202 &  337$|$399 &  -11$|$-34  &  \phantom{00}232$|$34696 &  \phantom{1}-24$|$294\phantom{-} &  -26$|$-38 \\  \hline \hline
\cat{OttExtra} \cite{ott:06prl}&&&&&&\\
\phantom{0}m15b6 & 1663$|$2165 &  -10$|$-35 &  -25$|$-53  &  \phantom{0}247$|$1212 &  -24$|$-41 &  -26$|$-52  &  \phantom{00}427$|$27296 &  \phantom{1}-22$|$220\phantom{-} &  -26$|$-41 \\ 
\phantom{0}s11WW & 1450$|$7562 &   \phantom{-}19$|$-25 &  -23$|$-51 & \phantom{0}272$|$1594 &  -24$|$-37 &  -26$|$-52  & \phantom{0}1056$|$39024 &  \phantom{1}-16$|$337\phantom{-} &  -25$|$-36 \\ 
\phantom{0}s25WW & \phantom{-0}7455$|$-50221 &   51$|$63 &  -23$|$-49  & $2\!\!\times\!\!10^5$$|$$4\!\!\times\!\!10^5$ & 2279$|$3801 &   \phantom{0}66$|$102  & $1\!\!\times\!\!10^5$$|$$1\!\!\times\!\!10^7$& \phantom{0}1075$|$$1\!\!\times\!\!10^5$&   \phantom{00}18$|$5236 \\  \hline \hline
\cat{MurRem} \cite{murphy:09}&&&&&&\\
\phantom{0}20\_3.4& 4253$|$9073 &   16$|$34 &  -24$|$-50  & 15668$|$22046 &  130$|$167 & -20$|$-43 &  -21$|$-32 &  -27$|$-53 &  -26$|$-52 \\ 
\phantom{0}12\_3.2& 1461$|$7459 &  -12$|$18\phantom{-} &  -25$|$-51  & \phantom{0}7905$|$14088 &   53$|$88 &  -23$|$-47  &  -25$|$-47 &  -27$|$-54 &  -26$|$-52 \\ 
\phantom{0}15\_3.2& \phantom{0}7450$|$10699 &   48$|$51 &  -23$|$-50  & 10862$|$17337 &  \phantom{0}82$|$120 & -22$|$-45 &  -25$|$-12 &  -27$|$-53 &  -26$|$-52 \\  \hline 
\cat{Yak} \cite{yakunin:10} &&&&&&\\
\phantom{0}s12\_matter  &  212$|$349 &  -24$|$-53 &  -26$|$-53  &  \phantom{0}777$|$1257 &  -19$|$-41 &  -26$|$-52  &  -24$|$-31 &  -27$|$-53 &  -26$|$-52 \\ 
\phantom{0}s15\_matter  &  194$|$504 &  -24$|$-52 &  -26$|$-54  & \phantom{0}902$|$2071 & -17$|$-33 & -26$|$-51 &  -25$|$5\phantom{0-} & -27$|$-53 &  -26$|$-52 \\ 
\phantom{0}s25\_matter  & 1099$|$1381 &  -15$|$-43 &  -25$|$-53  &  \phantom{0}726$|$2080 &  -19$|$-32 &  -26$|$-52  &  -16$|$-25 &  -26$|$-53 &  -26$|$-52 \\    \hline
\end{tabular}
\label{tab:extra}
\end{center}
\vspace*{-0.8cm}
\end{table*}

\subsection{Deciding Between Rotating Accretion-Induced Collapse and
Rotating Iron Core Collapse}

The waveforms of the \cat{Dim} catalog are representative of the GW signal
emitted by rotating collapse and bounce of iron cores of massive stars
with ZAMS masses $\gtrsim 8-10\,M_\odot$.  In the accretion-induced
collapse (AIC) of rapidly rotating O-Ne white dwarfs, very similar
dynamics occurs and the corresponding GW signals, as predicted by
Abdikamalov~\emph{et al.}~\cite{abdikamalov:10}, share many of the
basic features of the rotating iron core collapse and bounce waveforms
of, e.g., the \cat{Dim} catalog (see the discussion in Sec.~IV.C. of
\cite{abdikamalov:10}). Hence, it is interesting to see if our 
SMEE model selection algorithm can tell them apart.

We compute the PCs for the \cat{Abd} catalog in the same fashion as
done previously for the \cat{Dim}, \cat{Mur}, and \cat{Ott} catalogs
and inject all \cat{Abd} and \cat{Dim} waveforms into simulated
Advanced LIGO noise. SMEE is then run with 7 PCs to calculate $\log
B_{\tt AbdDim}$. The result is shown in Fig.~\ref{fig:AD} for source
distances of $10\,\mathrm{kpc}$ and $2\,\mathrm{kpc}$. Full numerical
results are available on-line \cite{smee1web}.

In spite of the strong general similarity of rotating iron core
collapse and rotating AIC waveforms, SMEE correctly identifies the
vast majority of injected waveforms as most likely being emitted by a
rotating iron core collapse or by rotating AIC.  However, for a source
at $10\,\mathrm{kpc}$ (left panel of Fig.~\ref{fig:AD}), $\sim$6\% of
the \cat{Dim} and $\sim$5\% of the \cat{Abd} are incorrectly identified
as belonging to the respective other catalog. For an additional 2\% of the
\cat{Dim} waveforms and 14\% of the \cat{Abd} waveforms, the evidence
is inconclusive.

At a source distance of  $2\,\mathrm{kpc}$ (right panel of Fig.~\ref{fig:AD}),
$88\%$ of the AIC (\cat{Abd}) and $93\%$ of the rotating core
collapse (\cat{Dim}) waveforms are correctly identified.  

If one placed trust in the reliability of less dominant and more
particular features of waveforms in the underlying catalogs, one could
use a larger number of PCs in the analysis. In order to study the
effect of using an increased number of PCs, we re-run the \cat{Abd}
vs.\ \cat{Dim} comparison with 14 PCs and find that the result is
significantly worse than with 7 PCs: $\sim$$61\%$ of the \cat{Abd}
waveforms and $\sim$$23\%$ of the \cat{Dim} catalog are now
incorrectly attributed to the respective other catalog at
$10\,\mathrm{kpc}$.  This counter intuitive and at first surprising
result is readily explained by the overall great similarity of the AIC
and iron core collapse waveforms and the nature of PCA and SMEE's
Bayesian model selection. The most robust features of each waveform
catalog are encapsulated in its first few PCs. The first \cat{Dim} and
\cat{Abd} PCs are indeed significantly different, but subsequent
\cat{Abd} and \cat{Dim} PCs exhibit rather similar secondary
features. Since each PC carries the same weight in SMEE's evidence
calculation, including a larger number of PCs dilutes SMEE's judgment
in this case and leads to the observed false identifications.

\begin{figure*}[tb]
\centering
\includegraphics[width=0.4\textwidth]{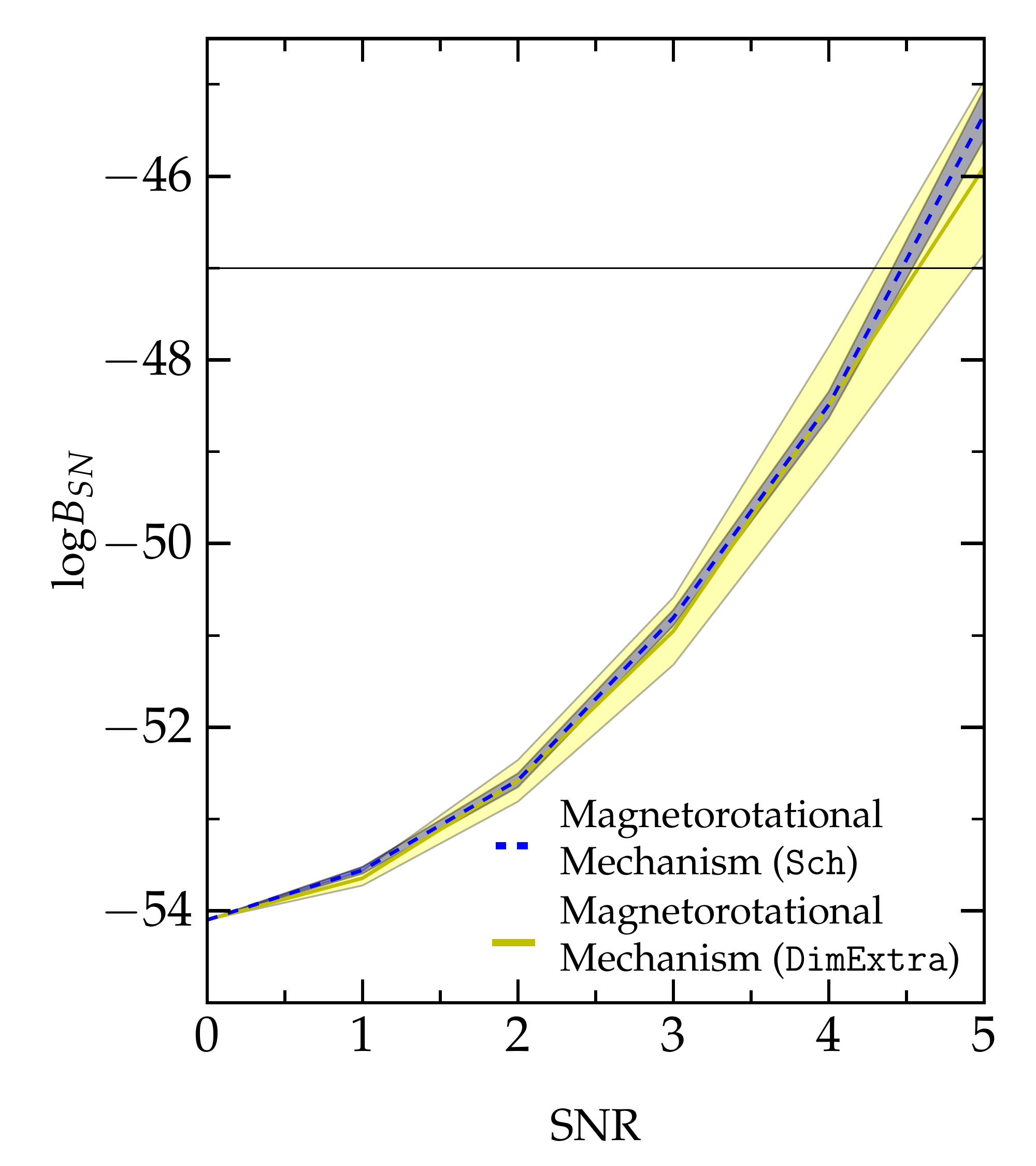}
\hspace*{0.35cm}
\includegraphics[width=0.4\textwidth]{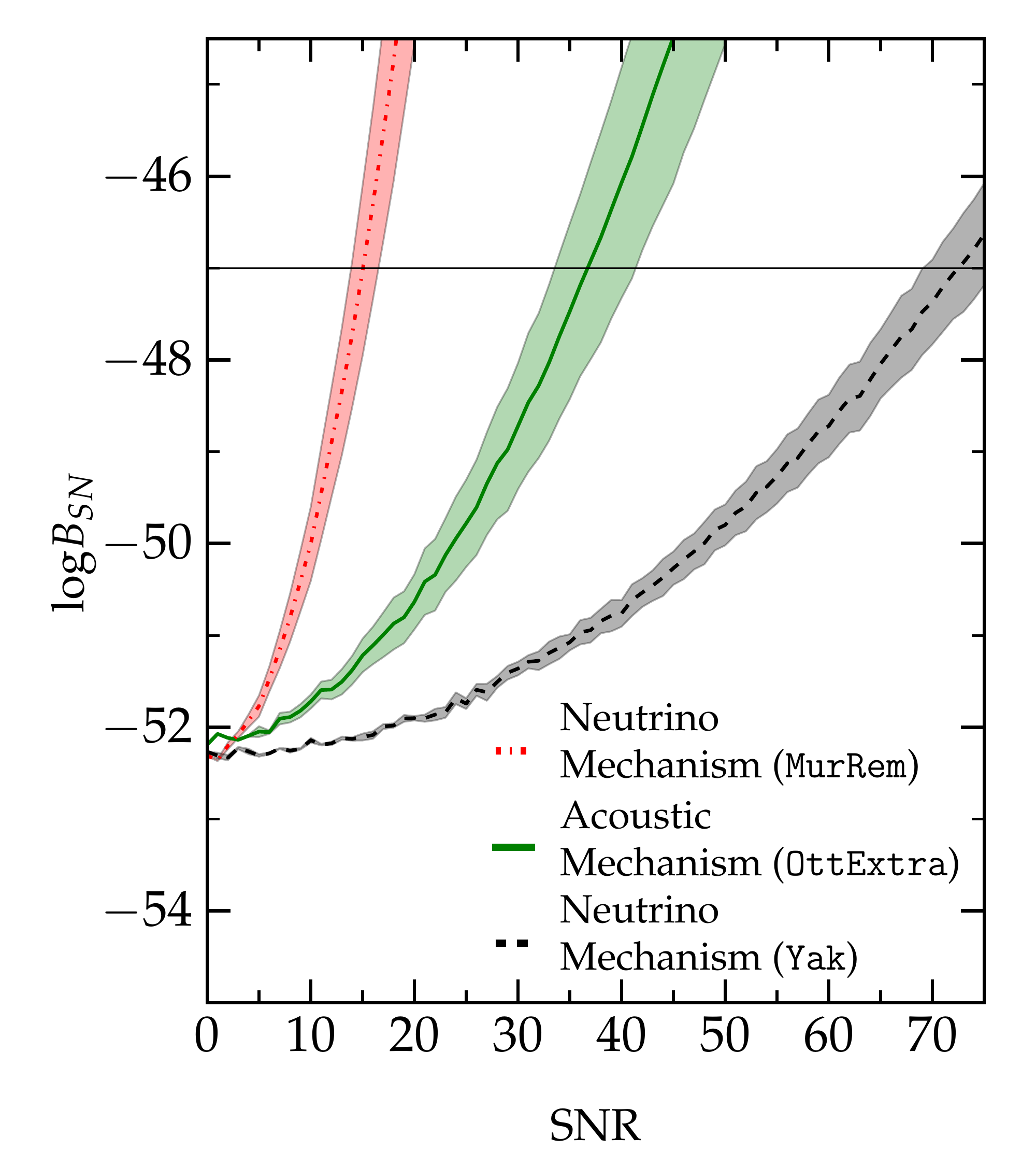}
\caption{Mean and $1$-$\sigma$ range of $\log{B_{SN}}$ as a function
  of signal-to-noise ratio SNR comparing signal with noise evidence.
  The horizontal lines mark the threshold values of $\log{B_{SN}}$
  above which we consider an injected waveform to be distinct from
  Gaussian noise. {\bf Left panel:} Results for the \cat{Sch} and
  \cat{DimExtra}. These two were both reconstructed with 7 \cat{Dim}
  PCs.  {\bf Right panel}: Results for the \cat{Yak}, \cat{MurRem} and
  \cat{OttExtra} waveforms as reconstructed with 7 \cat{Mur} for the
  first two and 7 \cat{Ott} PCs for the latter.  The \cat{Dim} PCs
  very efficiently reconstruct the \cat{Sch} and \cat{DimExtra}
  waveforms at moderate SNRs while the \cat{Yak} and \cat{OttExtra}
  require very high SNRs to be distinguished from noise by the
  \cat{Mur} and \cat{Ott} PCs, respectively.}
\label{fig:extraSNRs}
\end{figure*}

\subsection{Testing Robustness with Unknown Waveforms}

In the previous sections, we have demonstrated SMEE's ability to
identify an injected trial GW signal as belonging to a particular
physical model (i.e., emission mechanism and/or explosion mechanism).
In this, however, we have drawn the injected waveforms directly from
the catalogs used to generate the PCs. In other words, we have given
SMEE (limited) advance knowledge about the injected waveforms.

Here, we carry out a much more stringent test of SMEE's ability
to select between models of the core-collapse supernova mechanism
by injecting waveforms that were not employed in the initial PC generation
and/or stem from completely independent catalogs.

\subsubsection{Magnetorotational Mechanism}

For the magnetorotational mechanism, we employ three additional \cat{Dim}
waveforms (\cat{DimExtra}, Sec.~\ref{sec:mhdcats}) that were not included in the
calculation of the \cat{Dim} PCs. Furthermore, we inject waveforms from
rotating models of the \cat{Sch} catalog of Scheidegger~\emph{et~al.}\
\cite{scheidegger:10b,scheideggercat} (see
Sec.~\ref{sec:mhdcats}). The results of the $\log B_{SN}$ calculation
for the magnetorotational, neutrino, and acoustic mechanism signal
models are summarized in Tab.~\ref{tab:extra}. \cat{DimExtra} waveforms are
identified as being most consistent with the \cat{Dim} catalog and, hence,
the magnetorotational mechanism. This is true with high confidence
when 3 or 7 PCs are used and for all \cat{DimExtra} signals out to distances
$\gtrsim$$10\,\mathrm{kpc}$. 

The \cat{Sch} waveforms were generated with a completely different
numerical code and thus allow for a truly independent test of our
approach in SMEE. Also, unlike the \cat{Dim} waveforms, the \cat{Sch}
waveforms are based on 3D simulations. Hence, they are not linearly
polarized. For consistency with our current approach, we neglect
$h_\times$ and inject only $h_+$ as seen by an equatorial
observer. Results of SMEE $\log B_{SN}$ calculations for all injected
\cat{Sch} waveforms are summarized in Tab.~\ref{tab:extra}. SMEE
correctly identifies \emph{all} injected \cat{Sch} waveforms as
indicative of magnetorotational explosions at a source distance of
$2\,\mathrm{kpc}$. At $10\,\mathrm{kpc}$, still $91$\% of the injected
\cat{Sch} waveforms are attributed to the magnetorotational mechanism,
which is an indication of the robustness of the
GW signal associated with rapid rotation and magnetorotational
explosions.  The very few \cat{Sch} waveforms that SMEE is not able to
clearly associated with the magnetorotational mechanism have such weak
SNRs that they are more consistent with noise than with any of the
catalogs at $10\,\mathrm{kpc}$.

\subsubsection{Acoustic Mechanism}

We test SMEE's ability to identify core-collapse supernovae exploding
via the acoustic mechanism by injecting the three \cat{OttExtra}
waveforms (see Sec.~\ref{sec:acousticcats}). The results of this test
are again summarized in Tab.~\ref{tab:extra}. They suggest that the
a-priori unknown \cat{OttExtra} waveforms can be identified as
belonging to the acoustic mechanism out to $2\,\mathrm{kpc}$ with
great confidence when 7 PCs are used in the analysis. At
$10\,\mathrm{kpc}$, the waveforms are still correctly attributed to
the acoustic mechanism, but the evidence is much weaker in the 7-PC
case while the waveforms are more consistent with noise when the
analysis is performed with only 3 PCs. The \cat{OttExtra} 3 waveform,
which is clearly identified at $10\,\mathrm{kpc}$, has an extreme SNR
of $\sim$2530 at this distance, while the two other waveforms have
SNRs of $\sim$$50$. SMEE's difficulty is illustrated in the right
panel of Fig.~\ref{fig:extraSNRs}, which indicates that the
\cat{OttExtra} waveforms reach the threshold of $\log B_{SN} \ge -47$
only for SNRs $\gtrsim$$35$, whereas \cat{Ott} waveforms are
identified already at SNRs $\gtrsim$$4$, if the full set of 7 PCs is
used. This is a strong indication that the range of possible waveform
features associated with the acoustic mechanism is not efficiently
covered by the 7 PCs generated from the \cat{Ott} catalog.  This could
simply be attributed to the very small number of waveforms in this
catalog.  However, when studying the \cat{Ott} and \cat{OttExtra}
waveforms, one immediately notes that the time between the first peak
(associated with core bounce) and the second peak (the global maximum,
associated with the non-linear phase of the protoneutron star
pulsations) varies significantly between waveforms. Since we compute
PCs in the time domain, such large-scale features are imprinted onto
the PCs and make it difficult to identify waveforms whose two peaks
are separated by significantly different intervals. An alternative
method that may work much better for waveforms of this kind is to
compute PCs based on waveform power spectra, which would remove any
potentially problematic phase information.

\subsubsection{Neutrino Mechanism}

We test SMEE's ability to identify GW signals emitted by core-collapse
supernovae exploding via the neutrino mechanism in two ways. First, we
remove three randomly selected \cat{Mur} waveforms (\cat{Mur}
waveforms 20\_3.4, 12\_3.4, and 15\_3.2, labeling this set as
\cat{MurRem}) from the \cat{Mur} catalog, recompute the PCs without
these 3 waveforms, then run SMEE to compute $\log B_{SN}$ for the
three \cat{MurRem} waveforms. The results, listed in
Tab.~\ref{tab:extra}, show that SMEE is able to correctly
  identify \cat{MurRem} waveforms as GW signals consistent with the
  \cat{Mur} catalog with strong evidence out to a distance of
  $\sim$$2\,\mathrm{kpc}$ and even at $10\,\mathrm{kpc}$ two out of
  three \cat{MurRem} waveforms are picked out of the noise (though
  with relatively weak evidence). This is consistent with the overall
results for waveforms belonging to the \cat{Mur} catalog discussed in
Sec.~\ref{sec:mech}. 

However, a comparison of the right panel of Fig.~\ref{fig:extraSNRs}
with Fig.~\ref{fig:SNRs7PCs} shows that the \cat{MurRem} waveforms
require an SNR that is more than twice as high to reach values of
$\log B_{SN}$ at which we can consider them to be distinct from
Gaussian noise.  This is most likely due to the rather large diversity
of \cat{Mur} waveforms. Components of relevance to the \cat{MurRem}
waveforms are apparently not captured in the first 7 PCs when these
waveforms are not included in the PCA.

A yet more stringent test is enabled by the waveforms of the \cat{Yak}
catalog (see Sec.~\ref{sec:neutrinocats}) that were obtained with a
completely different numerical code. We inject the three available
\cat{Yak} waveforms into Advanced LIGO noise and run SMEE on them to
compute $\log B_{SN}$. We list the results in
Tab.~\ref{tab:extra}. SMEE correctly and clearly associates the
\cat{Yak} waveforms with the \cat{Mur} PCs at $0.2\,\mathrm{kpc}$.  At
$2\,\mathrm{kpc}$, the association is still possible, but at
$10\,\mathrm{kpc}$ the \cat{Yak} waveforms appear to be most
consistent with noise for SMEE.  The right panel of
Fig.~\ref{fig:extraSNRs} shows that the \cat{Yak} waveforms require an
SNR to be clearly associated with the neutrino mechanism that is
$\gtrsim$$7$ times higher than for \cat{MurRem} waveforms and more
than $\sim$$17$ times higher than for \cat{Mur} waveforms.  This
rather disappointing result can be explained as follows: While the
\cat{Yak} waveforms are qualitatively very similar to the \cat{Mur}
waveforms, they differ significantly in quantitative aspects. The
\cat{Yak} waveforms are generally only half as long
($\sim$$1\,\mathrm{s}$ for \cat{Mur} and $0.5\,\mathrm{s}$ for
\cat{Yak}, whose models explode much earlier than the \cat{Mur}
models).  Furthermore, the \cat{Yak} waveforms have considerably more
power at frequencies above $\sim$$800\,\mathrm{Hz}$ and their energy
spectra peak at $\sim$$1000\,\mathrm{Hz}$ while most of the emission
in the \cat{Mur} waveforms occurs at or below
$\sim$$400\,\mathrm{Hz}$. This may be due to the more simplified
treatment of gravity and neutrino microphysics and transport in the
study of Murphy~\emph{et al.}~\cite{murphy:09} underlying the
\cat{Mur} catalog compared to the work of Yakunin~\emph{et
  al.}~\cite{yakunin:10} that led to the \cat{Yak} catalog.

\section{Summary and Discussion}
\label{sec:summary}

In this article, we have described the Supernova Model Evidence
Extractor (SMEE), a novel Bayesian approach to inferring physical
information from observations of GW bursts emitted in
stellar collapse and core-collapse supernovae. SMEE decomposes
catalogs of simulated GW signals into their principle components and
employs the Nested Sampling algorithm to compute the evidence that a
trial signal injected into GW detector noise belongs to a given catalog.

It is evident that core-collapse supernovae are powered by the release
of gravitational energy in gravitational collapse, but the central
unsolved problem of core-collapse supernova theory is by what
mechanism this energy is transferred from the collapsed core to the
stellar mantle to drive the explosion. Simulations show that
multi-dimensional GW-emitting dynamics is a crucial ingredient to all
potential explosion mechanisms. Hence, GWs could be ideal probes for
the core-collapse supernova mechanism, provided that different
mechanism models have clearly distinct GW signatures. In this paper,
we have considered (\emph{i}) the neutrino mechanism, (\emph{ii}) the
magnetorotational mechanism, and (\emph{iii}) the acoustic mechanism
for core-collapse supernova explosions and have treated mechanism as
equally probable (i.e., having the same prior probability). The
primary and dominant multi-dimensional dynamics and GW emission
processes of these mechanisms are, ordered in the above order of
mechanisms, (\emph{i}) convection/turbulence and accretion
downstreams, (\emph{ii}) rapid rotation, and (\emph{iii}) protoneutron
star core pulsations, respectively.

Using GW signal catalogs based on simulations representative of these
three mechanisms, we have applied SMEE to infer the explosion
mechanism underlying trial waveforms injected into simulated Advanced
LIGO noise. Our results show that our Bayesian approach is capable of
identifying any of the considered explosion mechanisms with high
confidence for core collapse events occurring out to a distance of
$\sim$$2\,\mathrm{kpc}$ even with only rudimentary knowledge of the
precise shape of the expected signal.

Magnetorotational explosions, leading to particularly strong GW
signals with very robust common features, can be clearly identified
throughout the Milky Way ($D \gtrsim 10\,\mathrm{kpc}$). Our results
also suggest that it is possible to further distinguish between rapidly
rotating accretion-induced collapse of massive white dwarfs and
rapidly rotating iron core collapse for events occurring in the galaxy,
provided that the differences between the GW signals predicted by
current simulations of rotating accretion-induced collapse and
rotating iron core collapse are reliable.

GW signals emitted by neutrino-driven explosions have systematically
lower amplitudes and, hence, are harder to distinguish from detector
noise. Moreover, GW emission from convection/turbulence and accretion
downstreams has not been as extensively studied, the available number
of model waveforms is an order of magnitude smaller, and the currently
predicted GW signals are not as reliable as in the case of rapidly
rotating collapse.  This reduces the efficacy of the principal
component decomposition and, combined with the overall weakness of the
signals, limits SMEE's robust reach for the neutrino mechanism to
$\lesssim$$2\,\mathrm{kpc}$ for the currently available set of model
waveforms and Advanced LIGO in broadband configuration.

The GW signals from core-collapse supernovae driven by the acoustic
mechanism are strong and would likely be detectable by Advanced LIGO
throughout the Milky Way. However, the set of available model
waveforms for this signal type is very limited and individual
waveforms differ significantly at large scales in the time domain
while having similar frequency content. The inclusion of phase
information (by computing PCs in the time domain and using their full
complex Fourier transforms) in the current incarnation of SMEE
is sub-optimal for such signals and the PCs only inefficiently span the space
of possible waveforms. This makes it difficult for SMEE to clearly
identify the acoustic mechanism for core collapse events occurring at
distances significantly greater than $\sim$$2\,\mathrm{kpc}$.

This study is the first systematic attempt at inferring core-collapse
supernova physics from observations of GW bursts emitted by multiple
different underlying mechanisms. While this is a significant step
beyond previous work that focused only on GWs from rotating core
collapse \cite{summerscales:08,roever:09}, our present study still
suffers from a number of simplifying assumptions: We have considered
only a single detector, Gaussian noise, and linearly polarized,
optimally oriented GW emission. Real core-collapse supernovae will
emit in both GW polarizations and will have arbitrary, generally
non-optimal orientation with respect to observatories on
Earth. Advanced LIGO class GW detectors will operate as networks and
observations will be coincident between three or even four detectors,
and the noise backgrounds of these detectors will not be stationary and
Gaussian, but non-stationary and glitchy. We have also been agnostic
with regard to the prior probability of the three considered core-collapse 
supernova mechanisms. Input from astronomical observations and theory could
be used to generate (approximate) prior probabilities for each mechanism,
which could then be included in a future analysis.

Near future work on SMEE will be directed towards incorporating
detector networks, variations in detector configurations and
sensitivities, both GW polarizations, non-Gaussian noise, arbitrary
source--detector orientations, and improved principal component
analysis with and without reliance on the signal phase. However, even
with these improvements, successful extraction of physics from
core-collapse supernova GW signals will crucially depend on the
availability of extensive catalogs of reliable predictions for both
$h_+$ and $h_\times$. These must be provided by the core-collapse
supernova modeling community on the basis of full 3D simulations.


\acknowledgments

We thank the Basel, MPA Garching, ORNL, and Princeton core-collapse
supernova modeling groups for making their gravitational waveforms
publicly available.  We are happy to acknowledge helpful exchanges
with E.~Abdikamalov, A.~Burrows, Y.~Chen, D.~Chernoff, N.~Christensen,
T.~Loredo, J.~Murphy, J.~Nordhaus, L.~Santamaria, B.~Schutz,
M.~Vallisneri, A.~Weinstein, S.~Wesolowski and the core-collapse supernova working
group of the LIGO Scientific Collaboration and the Virgo
collaboration.

CDO is supported in part by the Sherman Fairchild Foundation and by
the National Science Foundation under grant number PHY-0904015.  Some
of the results presented in this article were obtained through
computations on the Caltech compute cluster ``Zwicky'' (NSF MRI award
No.\ PHY-0960291), on the NSF Teragrid under grant TG-PHY100033, on
machines of the Louisiana Optical Network Initiative under grant
loni\_numrel07, and at the National Energy Research Scientific
Computing Center (NERSC), which is supported by the Office of Science
of the US Department of Energy under contract DE-AC02-05CH11231.

ISH and JL gratefully acknowledge the support of the UK Science and Technology Facilities Council and the Scottish Universities Physics Alliance (SUPA).




\end{document}